\documentclass[referee]{aa} 
\usepackage{graphicx}
\usepackage{natbib}

\bibpunct{(}{)}{;}{a}{}{,}
\usepackage{txfonts}
\usepackage{longtable}
\usepackage{lscape}

\begin{document}

\title{Kinematics and dynamics of the luminous infrared galaxy pair NGC 5257/58 (Arp 240)\footnote{Reproduced with permission from Astronomy \& Astrophysics, \copyright ESO}}

\author{I. Fuentes-Carrera \inst{1}
\and M. Rosado \inst{2}
\and P. Amram \inst{3}
\and E. Laurikainen \inst{4} 
\and H. Salo \inst{4}
\and J. A. G\'omez-L\'opez \inst{3}
\and H. O. Castaneda \inst{1}
\and A. Bernal \inst{2}
\and C. Balkowski \inst{5}}

\offprints{I. Fuentes-Carrera}

\institute{Escuela Superior de F\'\i sica y Matem\'aticas, Instituto Polit\'ecnico Nacional (ESFM-IPN), U.P. Adolfo L\'opez Mateos, edificio 9, Zacatenco, 07730, Mexico City, Mexico  \\
\email{isaura@esfm.ipn.mx, isaura.fuentescarrera@gmail.com} 
 \and Instituto de Astronom\' \i a, Universidad Nacional Aut\'onoma de M\'exico (UNAM), Apdo. Postal 70-264, 04510, Mexico City, Mexico 
 \and Aix Marseille Universit\'e, CNRS, LAM (Laboratoire d'Astrophysique de Marseille), 13388,
Marseille, France 
 \and Department of Physical Sciences, Division of Astronomy, University of Oulu, FIN-90570, Oulu, Finland 
\and GEPI, Observatoire de Paris, CNRS, Universit\'e Paris Diderot, 5 Place Jules Janssen 92190 Meudon, France 
}

\date{Received . . . . . . . . . .  ; accepted . . . . . . . . . . }

\abstract
    {Encounters between galaxies modify their morphology, kinematics, and star formation history. The relation between these changes and external perturbations is not straightforward. The great number of parameters involved requires both the study of large samples and individual encounters where particular features, motions, and perturbations can be traced and analysed in detail.
}
    {We analysed the morphology, kinematics, and dynamics of two luminous infrared spiral galaxies of almost equal mass, NGC 5257 and NGC 5258, in which star formation is mostly confined to the spiral arms, in order to understand interactions between galaxies of equivalent masses and star-forming processes during the encounter.}
{Using scanning Fabry-Perot interferometry, we  studied the contribution of circular and non-circular motions  and the response of the ionized gas to external perturbations. We compared the kinematics with direct images of the pair and traced the star-forming processes and gravitational effects due to the presence of the other galaxy. The spectral energy distribution of each member of the pair was fitted. A mass model was fitted to the rotation curve of each galaxy. 
} 
{Large, non-circular motions detected in both galaxies are associated with a bar, spiral arms, and HII regions for the inner parts of the galaxies, and with the tidal interaction for the outer parts of the discs. Bifurcations in the rotation curves indicate that the galaxies have recently undergone their pericentric passage. The pattern speed of a perturbation of one of the galaxies is computed. Location of a possible corotation  seems to indicate that the gravitational response of the ionized gas in the outer parts of the disc is related to the regions where ongoing star formation is confined.
 The spectral energy distribution (SED) fit indicates a slightly different star formation history for each member of the pair.   For both galaxies, a pseudo-isothermal halo  better fits the global mass distribution.}
{}

\titlerunning{Interacting LIRG Arp 240}

\maketitle

\section{Introduction}
\label{intro}

Observations of distant galaxies suggest that mergers of galaxies played important roles in the formation of stars and galaxies in early epochs \citep[e.g.][]{2015A&A...575A..74S,2017MNRAS.465.1157R}. Studying the interacting galaxy population in the local universe is an important step towards a better understanding of the process of galaxy formation and evolution.  The way the interaction process triggers, sustains, or inhibits the formation of structure in spiral galaxies has been studied, from broad statistical studies to more detailed morphological and photometric analysis \citep[e.g.][]{2010A&A...509A..78D,2017MNRAS.464.1502M}. The kinematics and dynamics of a galaxy are also affected by encounters and interactions. Tidal features in interacting galaxies can give information on the stage of the encounter through kinematical effects on the baryonic matter, as well as the difference in rotation curves according to environment \citep[e.g.][]{2008A&A...484..299P}. Several attempts have been made to kinematically constraint the evolutionary scenario of interactions \citep[e.g.][]{2014A&A...566A..97J,2016A&A...591A..85B} and the formation of structures such as bars and tidal arms, amongst others \citep[e.g.][]{2010A&A...513A..43F,2015MNRAS.451.1307Z}. Early numerical simulations of interacting galaxies sought to reproduce the observed morphology; however, it has been shown that kinematical information imposes important restrictions on any interaction model \citep[e.g.][]{1993ApJ...410..586S,2000MNRAS.319..377S,2000MNRAS.319..393S,2009ApJ...702..392G,2014ApJ...787...39G}. 

The correlation between enhanced star formation (SF) and external perturbations is not straightforward \citep[e.g.][]{2015MNRAS.449.3719S,2015MNRAS.452..616D,2017MNRAS.467...27K}. In most studies using the partly or completely far-infrared (FIR) flux to estimate star formation rates (SFR) in galaxies, a clear enhancement is found in interacting systems \citep[e.g.][]{2013MNRAS.430.3128E}. In 1999, Gao and Solomon showed that the star formation efficiency (SFE) increases as separation between merger nuclei  decreases, suggesting that the starburst phase  occurs at a late stage of merging  and is confined to the central kiloparsec \citep{1999ApJ...512L..99G}. This effect has been shown for example in the work of  \citet{2017ApJ...844...96Y}. Optical, near-infrared (NIR), mid-infrared (MIR), and radio continuum studies also imply enhanced central SF in interacting galaxies \citep[e.g.][]{ 2007AJ....133..791S,2011MNRAS.412..591P,2013MNRAS.430.3128E,2015ApJS..218....6B}.  However, when unbiased samples of interacting and non-interacting galaxies are compared, the optical studies show only a moderate excess of SF concentrated in the centres of the interacting galaxies \citep[e.g.][]{2003A&A...405...31B,2015MNRAS.454.1742K}. 

Regarding the role of dark matter (DM) in the dynamics and evolution of interacting galaxies, the picture is less clear. In the standard picture of hierarchical structure formation, larger DM halos are built up through the accretion and merger of smaller halos. Most of the work on the dependence of DM halos on environment has been done using statistical results from large numerical simulations \citep[e.g.][]{2012MNRAS.419.2133H}. Observationally, the structure and distribution of DM halos has been found to be  related to the environment of their galaxies  \citep[e.g.][]{2015MNRAS.447..298H}. Following the hierarchical scenario and observations, isolated galaxies at a given cosmic epoch have formed pairs, triplets, and groups at later stages of evolution; some of those structures join to form galaxy clusters. From field galaxies to galaxies in rich clusters, galaxies should merge together to eventually form larger structures; dark halos being much more extended than their embedding galaxies, at some point the mass in the galaxies’ individual halos must migrate into a shared halo.
Numerical simulations indicate that this process starts when the distance between galaxies reaches the halo virial radius \citep{2004ApJ...608..663K,2008MNRAS.391.1685S}. However, this is not directly observed and the details of the redistribution of DM during these encounters is poorly understood. Galaxy interactions are still the best option to trace and probe the matter redistribution during such events. Those processes being dynamical, the morphological information on the encounter, even if instructive in the case of for example tidal tails and plume formation, is not sufficient to characterise the interaction. Kinematical data are needed to understand the dynamics of the system and its three-dimensional (3D) geometry.

We present and analyse scanning Fabry-Perot (FP) interferometric observations of the interacting galaxy pair Arp 240 (KPG 389), composed of two luminous infrared galaxies (LIRGs), NGC 5257 and NGC 5258. Section \ref{obs} presents the observations and data reduction. Section \ref{arp240_desc} introduces the  pair of galaxies. In Sect. \ref{kin} we discuss the kinematic information derived from the observations along with direct images of the pair in different wavelengths in order to compare between kinematical and morphological features. Section \ref{dyns} presents dynamical results, mass estimates, and mass distribution. In Sect. \ref{sed} we show the fit of the spectral energy distribution of both galaxies with estimates of their star formation history. A discussion is presented in Sect. \ref{disc} and conclusions are presented in Sect. \ref{ccl}.
The cosmological model adopted  in  this work  has $H_{0}$=67.8 \ km s$^{-1}$ Mpc$^{-1}$, $\Omega_{m}=$0.308, and $\Omega_{\lambda}=$0.692 \citep{2016A&A...594A..13P}.

\section{Observations and data reductions}
\label{obs}

Observations of Arp 240 were done at the 2.1 m telescope of the Observatorio Astron\'omico Nacional in San Pedro M\'artir, Mexico (OAN-SPM) using the scanning Fabry-Perot interferometer PUMA \citep{1995RMxAC...3..263R}. PUMA is a focal reducer used to make direct imagery and Fabry-Perot (FP) interferometry of extended emission sources. Instrumental specifications are presented in Table \ref{inst}. A $1024 \ \times \ 1024$ Tektronix CCD detector was used with a sampling of 0.58 $ \arcsec /$  px per axis. A physical $2 \times 2 $ binning in both spatial dimensions was done by the CCD controller in order to enhance the signal-to-noise ratio (SNR) in the readout noise per pixel by a factor of four. The final spatial sampling thus equals to $1.16 \ \arcsec /$ px . In order to isolate the redshifted H$\alpha$ emission of the galaxies, an interference filter centred at  $ 6720 \ \AA $ with a full width at half maximum (FWHM) of $ 20 \ \AA $ was used. The interference filter having a Gaussian-like shape allows us to pass wavelengths ranging from 6706 $\AA$ to 6734 $\AA$.  To average the sky variations during the exposure, we got two data cubes with an exposure time of 48 minutes each (60 s per channel times 48 channels), which were co-added leading to a total exposure time of 96 minutes. For the calibration we used a Ne lamp since the  $6717 \ \AA $ Ne line was close to the redshifted H$\alpha$ galaxy wavelength. Two calibration cubes were obtained at the beginning and at the end of the exposure in order to account for metrology.
Observational parameters are shown in Table \ref{inst}.
 
Data reduction and analysis were done using mainly the ADHOCw software developed by J. Boulesteix \citep{1989A&AS...81...59A} and the CIGALE software \citep{1993A&A...280..365L}. Standard corrections (cosmic rays removal, bias subtraction, and  flat-fielding) were done on each data cube before co-addition. Once co-added, using the calibration data cube, the resulting science data cube was calibrated in wavelength, meaning that each profile for each pixel has the same wavelength origin (and range). Then the night sky continuum and OH night sky lines were subtracted. A spectral Gaussian smoothing (with a velocity dispersion equal to two channels) has been applied. 
 
Details on the derivation of continuum, monochromatic, and velocity maps from scanning FP observations are given in \citet{2004A&A...415..451F}. In order to get a sufficient signal-to-noise ratio on the outer parts of each galaxy, we performed three spatial Gaussian smoothings ($\sigma = 2.32\arcsec, \ 3.48\arcsec, \ 4.64\arcsec$) on the resulting calibrated cube, so that a variable resolution radial velocity map was built using less spatially and spectrally smoothed pixels for regions with an originally higher signal-to-noise ratio  and more spatially and spectrally smoothed pixels for the outer parts of the galaxy.

\begin{flushleft}
\begin{table}
\centering \caption{Instrumental and observational parameters.}
\label{inst}
 \begin{tabular} {l l l }
\hline\hline
 \noalign{\smallskip}
Parameter   & &   Value     \\
 \noalign{\smallskip}
\hline
 \noalign{\smallskip}
Telescope & & 2.1 m  (OAN-SPM)    \\
Instrument & &  PUMA  \\
Scanning F-P interferometer & & ET-50 (Queensgate) \\
Interference order                                 & & 330 at 6562.78 \AA    \\
Free spectral range, $FSR$ \  (\AA \ / $km \ s^{-1}$)  & & 19.95 / 912.0       \\ 
Interferometer finesse, $\mathcal F$               & & 24 \\
Number of scanning channels  & & 48 \\
Spectral sampling at H$\alpha$ \   (\AA \ / $km \ s^{-1}$) & &  0.42 / 19.0      \\ 
Spectral resolution, R & & 6720/0.83     \\
Spectral element  (\AA \ / $km \ s^{-1}$)  & &   0.83 / 38.5     \\
Interference filter (\AA)  & &  6720 \\
$FWHM$ of filter  (\AA)    & & 20    \\
Detector & & Tektronix CCD      \\
Detector size (px x px) & & 1024 x 1024 \\
Image scale ( \arcsec per px ) & & 0.58   \\
Electronic binning & &  2     \\
Image scale after binning    ( \arcsec per px ) & & 1.16       \\
Mean seeing during observations (\arcsec) & & 1.5  \\
Total exposure time  (min)        & & 96       \\
Total exposure time per cube (min) & & 48       \\
  Number of science cubes & & 2     \\
Exposure time per channel (sec) & & 60       \\
Exposure time for direct image  (sec) & & 120       \\
Calibration line (\AA) & & 6717  \ (Ne lamp)       \\
\noalign{\smallskip} \hline
\end{tabular}
\end{table}
\end{flushleft}

\section{Arp 240: NGC 5257 and NGC 5258}
\label{arp240_desc}

\subsection{The pair of galaxies}
\label{arp240_pair}

\begin{figure}
\centering
\includegraphics[width=0.8\textwidth]{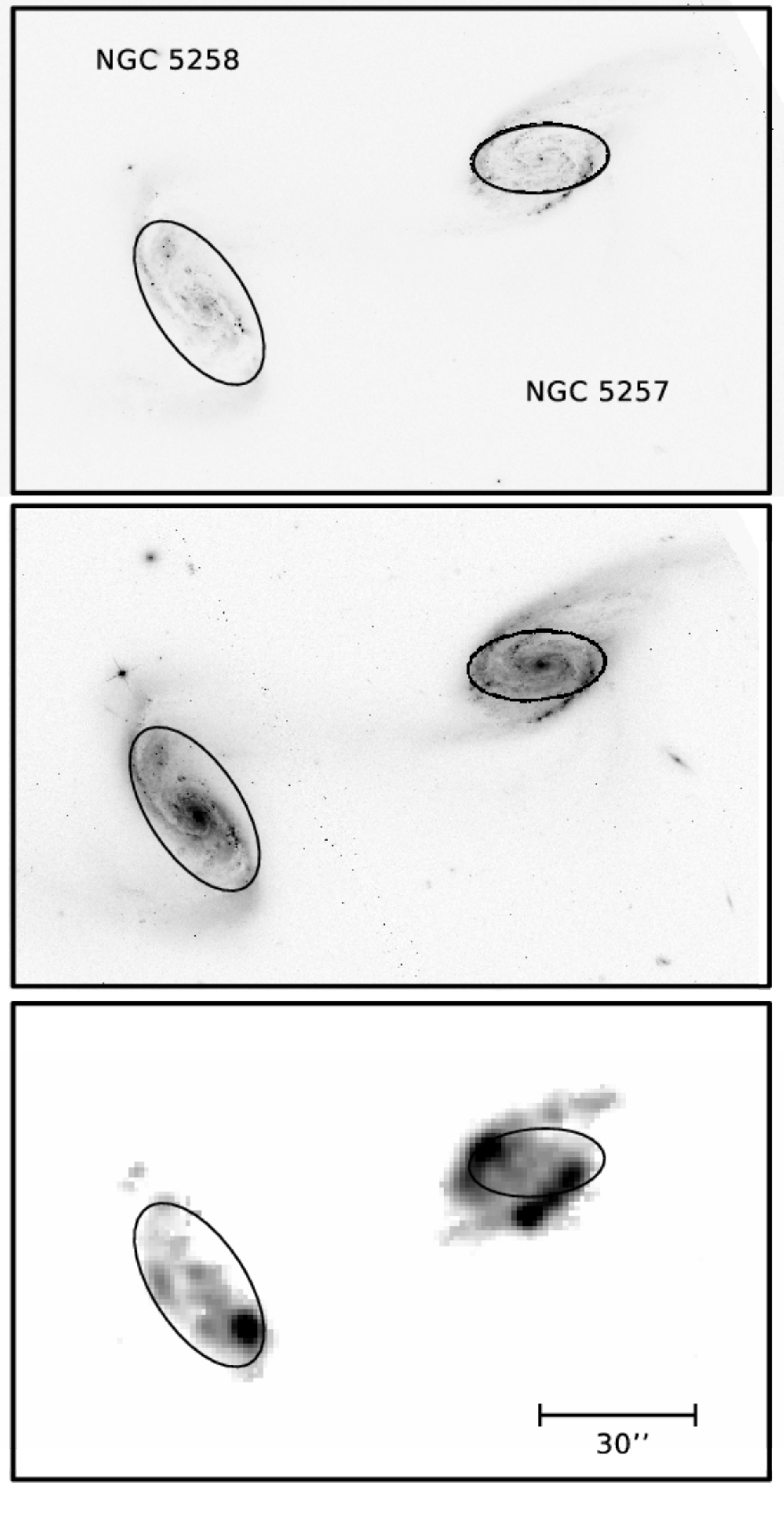}
\caption{ Top panel: High spatial resolution HST image of Arp 240 taken with the Advanced Camera for Surveys (ACS) in its Wide Field Channel (WFC) in the  F435W  filter (between U- and B-bands).
Middle panel: HST-ACS WFC image in the  F814W filter (close to the I-band). 
Bottom panel: Monochromatic H$\alpha$ (continuum subtracted) image  of the pair obtained from the scanning Fabry-Perot interferometer PUMA data cubes. 
Ellipses indicate the location of the bifurcation radius, $R_{bif}$ , in the rotation curve of NGC 5257 and NGC 5258, respectively. These are discussed in Sect. \ref{rot_curves}.
HST images taken from the HST archive, proposal 10592 by Aaron Evans.
NGC 5257 is the galaxy on the right of the image, and NGC 5258 is the galaxy on the left. North is to the top, east is to the left.
} 
\label{arp240_img}
\end{figure}

Arp 240 is an interacting galaxy pair consisting of two spiral galaxies: NGC 5257 and NGC 5258 (Figure \ref{arp240_img}).  Using the line-of-sight (LOS) radial velocities of the galaxies, corrected for the Local Group infall onto Virgo (6798 \ km \ s$^{-1}$ and 6757 \ km \ s$^{-1}$, for NGC 5257 and NGC 5258, respectively\footnote{Values taken from the LEDA database.}), the galaxy pair is located at a distance of $\sim$100.0 \ Mpc. This implies linear diameters of 51.7 kpc (106.6 $\arcsec$) and 48.3 kpc (99.6 $\arcsec$), for NGC 5257 and NGC 5258, respectively, considering the $D_{25}$ values by \citet{1991rc3..book.....D}, and a projected separation between the two galaxies of 37.8 kpc (78$\arcsec$). 
The general parameters of both galaxies are presented in Table  \ref{glxs_parameters}.

The pair belongs to the sample of bright Infrared Astronomical Satellite (IRAS) galaxies \citep{1989AJ.....98..766S}. The total infrared luminosity (L$_{IR}$) of the system is $2.8 \times 10^{11} L_\odot$, and the total FIR luminosity equals $2.02 \times 10^{11} \ L_\odot$ \citep{1993PASJ...45...43S}, with both galaxies contributing more or less equally to it. From their flux ratios, \citet{1995ApJS..100..325L}  classified both galaxies as starbursts. Simulations of this pair have been done by \citet{2016MNRAS.459..720H}  seeking to reconstruct the disturbed morphologies of interacting galaxies with the use of a restricted three-body simulation code, however  no kinematical restrictions were considered for the modelling.
The top panel of Fig.  \ref{arp240_img} shows  the high spatial resolution Hubble Space Telescope (HST) image\footnote{HST archive, proposal 10592 by Aaron Evans.} of the pair taken with the Advanced Camera for Surveys (ACS) in its Wide Field Channel (WFC) using the F435W filter (between the U- and B- bands). The middle panel of Fig. \ref{arp240_img} shows the HST-ACS F814W  (close to the I-band)  image of the pair.  Figure \ref{arp240_ultracont} shows the high-contrast B-band image of Arp 240 displaying the tidal features of each galaxy.

\begin{table*}

\centering 

\caption{General parameters of NGC 5257 and NGC 5258.}
\label{glxs_parameters}

\begin{tabular}{l l l l l l l}

\hline
\hline
\noalign{\smallskip}
Parameter              &  &    NGC 5257  & \ \ \ \ \ \ \ \ \ \ \ &  NGC 5258  & & Reference  \\
\noalign{\smallskip}
\hline
\noalign{\smallskip}
Coordinates (J2000)\tablefootmark{a} & & $\alpha = 13h \ 39m \ 52.91s  $               & & $\alpha = 13h \ 39m \ 57.70s $                  & & ...  \\
                                     & & $\delta = +00^\circ \ 50\arcmin \ 24.5\arcsec$ & & $\delta = +00^\circ \ 49\arcmin \ 51.1\arcsec $  & & ...  \\

Morphological type  & & SABb\tablefootmark{a}                  & &  SBb\tablefootmark{a}                   & &  ...   \\

                    & & SAB(s)b pec HII LIRG\tablefootmark{b}  & &  SA(s)b:pec; HII LINER\tablefootmark{b} & &  ...    \\

                    & & SA(rs)b pec                            & &  SABb pec                               & &   1   \\

Heliocentric systemic        & &  $ 6800 \pm 4$ \tablefootmark{a,c}  &  &  $ 6782 \pm 9 $\tablefootmark{a,c}   & &  ...   \\

velocity  ($km \ s^{-1}$)     & &  $ 6764$\tablefootmark{d}           &  &  $ 6765 $\tablefootmark{d}           & &  2   \\

Redshift\tablefootmark{b}    & & $0.022676$                 & & $0.022539$                 & & ... \\

Distance (Mpc)               & & $98.0$                    &  & $97.4$                     & &  3   \\

                             & & $100.3$\tablefootmark{e}  &  & $99.7$\tablefootmark{e}    & &  ...  \\

Morphological position angle\tablefootmark{a} ($^\circ$)    & & $ 83.1 $  & &  $178$   & &  ...  \\

Morphological inclination\tablefootmark{a}  ($^\circ$)      & & $ 58$     & &  $ 34$   & &  ...  \\

 $m_{B}$ (mag)                 & &  $11.23$             & &  $14.05$           & &  1   \\

 $m_{R}$ (mag)                 & &  $10.28$             & &  $13.18$           & &  1 \\
 
 $m^{\mathrm{f}}_{Kext}$ (mag)   & &   $10.084 \pm 0.047$  & & $9.962 \pm 0.039$  & & 4  \\

$(B-I)$                                     & & $1.59$   & & $1.47$   & &  1 \\

$B^{\mathrm{0}}_{T}$\tablefootmark{g}        & & $11.03$  & & $13.90$  & &  1 \\

$(B-V)^{\mathrm{0}}_{T}$ \ \tablefootmark{h}   & & $0.51$   & & $0.42$   & &  1 \\

$R_{25} = D_{25}/2$ in arcsec (kpc)              & &  $53.3 \ (25.8) $   & & $49.8 \ (24.1)$    & &  5  \\

$R_{K20}$\tablefootmark{f} in arcsec (kpc)       & &  $29.9 \ (14.49) $  & & $37.6 \ (18.22)$  & &  4   \\

$R_{H\alpha} $\tablefootmark{i} in arcsec (kpc)  & &  $ 17.98 \ (8.71)$  & &  $18.04 \ (8.75)$  & &  ... \\

Mean surface brightness               & &          & &           & &     \\

within 25 isophote $(mag/arcsec^{-2})$ & & $22.74$  & & $23.26$  & &   1  \\

$L_{H\alpha}$ $(10^{6} L_\odot)$   & &  $5$     & &  $6$                 & &  6  \\

$L_{IR}$ $(10^{11} L_\odot)$      & &  $1.7$   & &  for the whole pair  & &  7  \\

                                & &  $2.7$   & &  for the whole pair  & &  8  \\

 $L_{FIR}$ $(10^{11} L_\odot)$    & &  $1.7$   & &    ...                & &  9 \\

                               & &  $2.0$   & &   for the whole pair   & &  6   \\

                               & &  $1.5$   & &   for the whole pair   & &  8    \\

                               & &  $2.0$   & &   for the whole pair   & &  10   \\

$L_{IR} / L_B$                   & & $12.0$  & & for the whole pair      & & 7    \\

Maximum rotation velocity ($km \ s^{-1}$)  & &  $ 248 \pm 10$\tablefootmark{a}  & &  $ 387 \pm 19$\tablefootmark{a}   & &   ...  \\

                                           & & $ 270$\tablefootmark{j}         & &  $ 250$\tablefootmark{j}          & &  2   \\

Dynamical mass  ($10^{11} \ M_\odot$)       & &  $ 0.28$\tablefootmark{k}        & &  $ 1.10$\tablefootmark{k}         & &  2  \\

Mass in HI (in $ 10^{10} \ M_\odot$) & &  $1.2$  & & $0.98$                & &  6   \\

                                   & &  $3.39$ & & for the whole pair    & &   2   \\

Mass in H$_2$  (in $ 10^{10} \ M_\odot$)  & &  $1.32$ & & $2.40$   & &  2           \\

SFR\tablefootmark{l}  ($M_\odot \ yr^{-1}$)   & & $27.8$    &  & $24.9$ & &   2,11  \\

\noalign{\smallskip}

\hline

\end{tabular}

\tablebib{ (1)~\citet{2001A&A...379...54H}; (2) \citet{2005ApJS..158....1I}; (3) \citet{2000ApJ...529..786M}; (4) \citep{2006AJ....131.1163S}; (5) \citet{1991rc3..book.....D}; (6) \citet{1993PASJ...45...43S}; (7) \citet{1998AJ....115..938B}; (8) \citet{2002ApJS..143...47D}; (9) \citet{2004A&A...422..941C}; (10) \citet{2000MNRAS.318..124G}; (11) \citep{2001ApJ...554..803Y}.
}

\tablefoot{
\tablefoottext{a} {HyperLEDA.}
\tablefoottext{b} {NED.}
\tablefoottext{c} {From radio observations.}
\tablefoottext{d} {Median velocity of the CO(1-0) spectrum.}
\tablefoottext{e} {From redshift value.}
\tablefoottext{f} {2MASS database $k\_m\_k20fe$ value.} 
\tablefoottext{g} {Total corrected magnitude in the RC3 system.}
\tablefoottext{h} {Total corrected colour index in the RC3 system.}
\tablefoottext{i} {This work.}
\tablefoottext{j} {From rotation curve with no inclination correction.} 
\tablefoottext{k} {Derived from the maximum radial extent of the CO(1-0) emission: $4.7 \ kpc$ ($ 0.18 \ R_{25}$) for NGC 5257, $12 \ kpc$ ($0.50 \ R_{25}$) for NGC 5258, and the observed rotational velocity divided by $ \sin 90$.}
\tablefoottext{l} {From $L_{1.4 \ GHz.}$ } 
}

\end{table*}



\subsection{NGC 5257}
\label{arp240_n5257}

NGC 5257 is classified as an SABb galaxy according to the HyperLEDA\footnote{HyperLEDA database http://leda.univ-lyon1.fr/.} database and as an SAB(rs)b pec HII LIRG galaxy in the NED\footnote{The NASA/IPAC Extragalactic Database (NED) is operated by the Jet Propulsion Laboratory, California Institute of Technology, under contract with the National Aeronautics and Space Administration.} database. General parameters of this galaxy are presented in Table \ref{glxs_parameters}. 
Top panel of Fig.  \ref{arp240_img} shows the western spiral arm  is outlined by small bright ``pearl-like'' regions.  In the middle panel of Fig. \ref{arp240_img} a small bulge can be seen embedded in a faint bar  $\sim4\arcsec$ in length. Inner spiral arms are well-traced coming out of the bar.  The external parts of both arms seem to be tidally stretched. 
From near-infrared (NIR) and H$\alpha$ observations of this galaxy, \citet{1998AJ....115..938B} found high levels of SF activity in the western arm of NGC 5257, while the nucleus shows a low degree of SF activity.  A prominent stellar bar ($6.5 \arcsec = 3.1$ kpc, in length) can be seen in the $K$-band image by \citet{1998AJ....115..938B}. \citet{2005ApJS..158....1I} studied the HI, radio continuum, and CO(1-0) emissions of this galaxy using the Very Large Array (VLA) and the Owens Valley Radio Observatory (OVRO) millimeter array, respectively. Figure 3 of their work shows a relatively undisturbed HI distribution for NGC 5257 but exhibits a CO(1-0) concentration near the nucleus extending to the north-east and to the south-west, forming a bar 10 kpc long. Using the radio continuum fluxes,  those authors estimated an SFR of $27.8 \ \ M_\odot \ yr^{-1}$. Figure \ref{arp240_ultracont} shows two prominent tidal arms for this galaxy. The western arm is elongated becoming an apparent tidal bridge between both galaxies, while the eastern arm traces the beginning of a tidal tail.

\subsection{NGC 5258}
\label{arp240_n5258}

\begin{figure}
\centering
\includegraphics[width=0.8\textwidth]{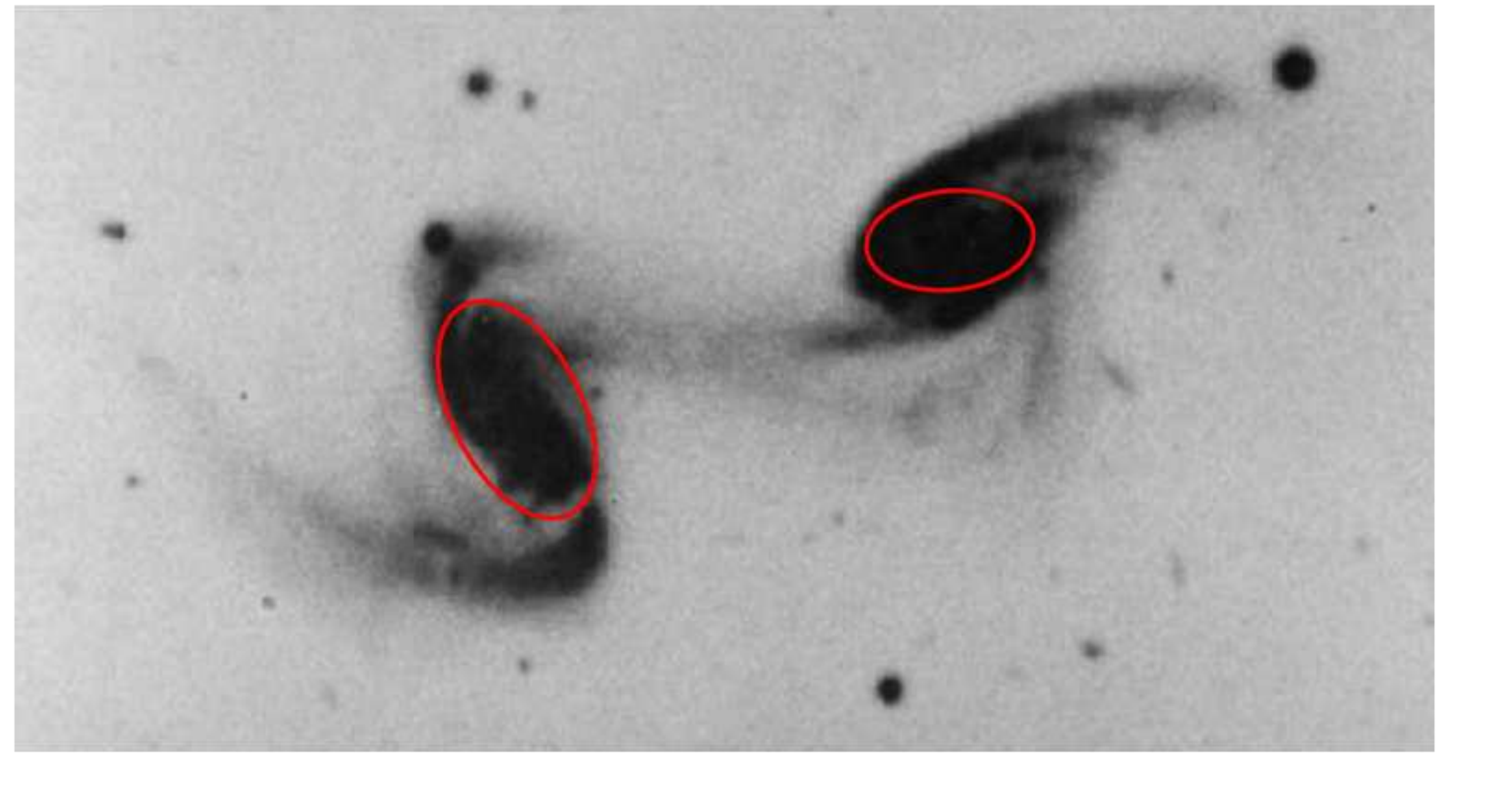}
\caption{High-contrast B image of Arp 240 showing the extension of the tidal features of each galaxy, taken from the "Catalog of Peculiar Galaxies'' \citep{1966ApJS...14....1A}.  
The ellipses indicate the location of the bifurcation radius, $R_{bif}$ , in the rotation curve of NGC 5257 and NGC 5258, respectively. These are discussed in Sect. \ref{rot_curves}.
}
\label{arp240_ultracont}
\end{figure}

NGC 5258 is classified as an SBb galaxy in the HyperLEDA database and as an SA(s)b?pec HII low-ionization nuclear emission-line region (LINER) in the NED database. In their detailed analysis of the optical surface photometry of (S+S) galaxies, \cite{2001A&A...379...54H} classified NGC 5258 as SABb pec. General parameters of this galaxy are presented in Table \ref{glxs_parameters}. The HST-ACS {\it F435W} image of this galaxy (top panel of Fig.  \ref{arp240_img}) shows the bulge is surrounded by a ring-like structure. The northern arm of the galaxy bifurcates into a northern and western arm. This bifurcation is clearly marked by dust. This is also seen in the HST-ACS {\it F814W} image (middle panel of Fig.  \ref{arp240_img}). A small bar almost aligned with the position angle, P.A., of the galaxy is detected in this filter. Dust lanes trace part of the spiral arms.  Important emission is seen along the southern arm in what seems to be a series of stellar clusters. These clusters are also seen in the $F413W$ image. In their optical and NIR photometry analysis of interacting galaxies, \citet{1998AJ....115..938B} found that the H$\alpha$ emission map of this galaxy shows star-forming activity in the region south-west of the nucleus, coinciding with the H$\alpha$ emission peak detected by \citet{2002ApJS..143...47D}, while the nucleus itself shows low star-forming activity.  The VLA images by \citet{2005ApJS..158....1I} show an HI tail $\sim125 \arcsec $  (60 kpc) long, extending beyond the optical tail. The CO(1-0) emission detected by \citet{2005ApJS..158....1I}  closely traces the optical morphology. Figure \ref{arp240_ultracont} shows that the northern extension of the northern arm of NGC 5258 appears to be truncated, while the western extension points towards the companion and seems to join it through a tidal bridge. The southern arm of the galaxy broadens and becomes an extended tidal tail.

\section{Kinematic results}
\label{kin}

\subsection{H$\alpha$ Image}
\label{Halpha}

The bottom panel of Fig. \ref{arp240_img} shows the H$\alpha$ line intensity image of the pair derived from PUMA observations by integrating the monochromatic H$\alpha$ profile in each pixel of the data cube. For NGC 5257, HII regions outline both arms, especially the western arm. For the latter, the same three HII regions detected in the H$\alpha$ images by \citet{1998AJ....115..938B}   and \citet{2002ApJS..143...47D}  are seen. A prominent HII region is observed in the middle part of the  eastern arm. The centre of the galaxy is devoid of HII regions. H$\alpha$ emission is seen along the tidal tail and  the beginning of the bridge between the two galaxies. For NGC 5258, a prominent HII region is seen in the south-western side of the galaxy, and  much fainter HII regions are seen in the beginning of the northern arm.  These regions are also seen in the H$\alpha$ images by \citet{1998AJ....115..938B}    and \citet{2002ApJS..143...47D}. The tidal bridge visible in the high-contrast B-band image is not detected in H$\alpha$.

\subsection{Velocity Fields}
\label{VFs}

Table  \ref{glxs_parameters} indicates that both galaxies are classified as barred or non-barred galaxies depending on the authors, from SA to SAB for NGC 5257 and from SA to SB for NGC 5258. Visual inspection of the HST {\it F814W} image makes plausible the existence of small bars of equal size in both galaxies,  but slightly stronger for NGC 5258 than for NGC 5257. On the other hand, the bar is almost aligned with the major axis of NGC 5258, while it makes roughly an angle of $\pi$/4 in the case of NGC 5257.

\begin{figure*}
  \centering
\includegraphics[width=0.8\textwidth]{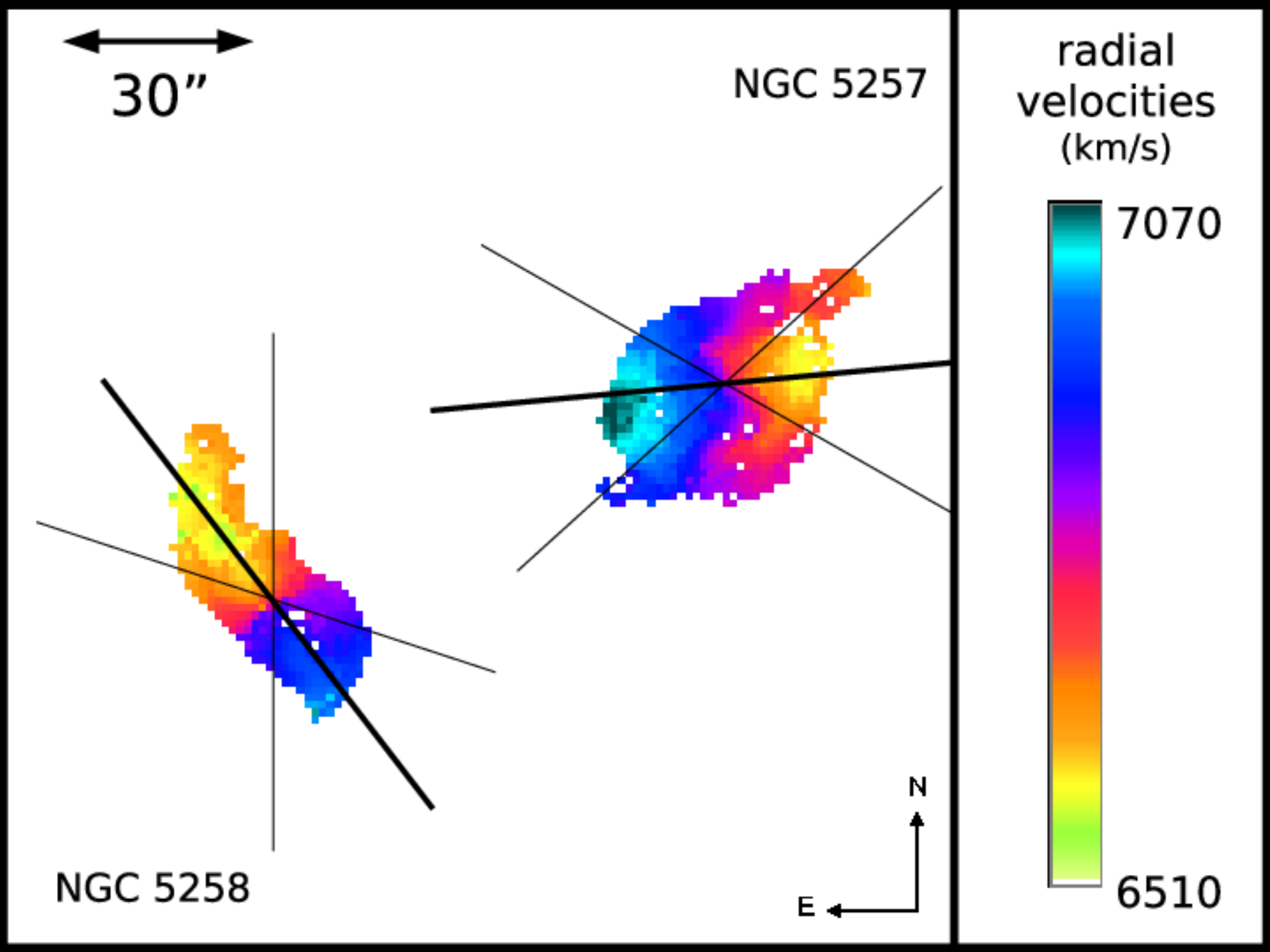}
\includegraphics[width=0.8\textwidth] {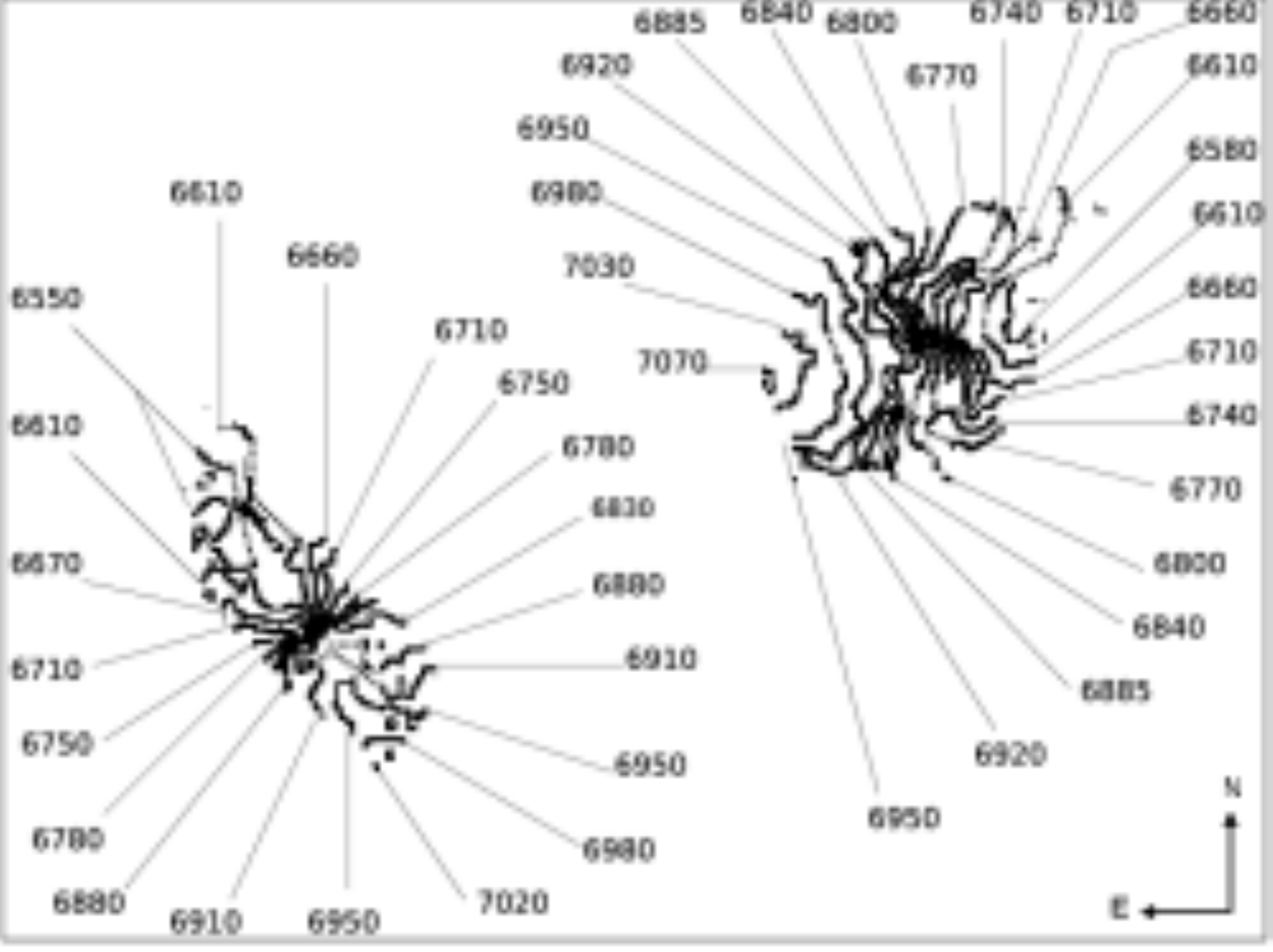}
\caption{ Top: Velocity field of Arp 240 (NGC 5257/58).
The solid thick line indicates each galaxy's kinematic position angle (P.A.),
the solid thin lines indicate the angular sectors from both sides of the
major axis considered for the computation of each galaxy's rotation curve. 
Bottom:
Isovelocities plot for Arp 240 (NGC 5257/58). } \label{arp240_vf}
\end{figure*}

The top panel of Fig. \ref{arp240_vf} shows the velocity field of both galaxies; the right panel of the same figure displays the corresponding isovelocities. The innermost parts of NGC 5257 exhibit isovelocities almost parallel to each other and that are not perpendicular to the major axis of the galaxy within a region that has an extension of  $\sim7.8 \arcsec $ (3.8 kpc). This is related to the presence of a stellar bar oriented between the major and the minor axis of the galaxy as can be seen on the HST images and on the velocity field itself. For NGC 5258, the isophotes along the minor axis are nearly straight with velocities around 6780 \ km \ s$^{-1}$. A little further away from the centre ($\sim$ 2.4 kpc, 5 $\arcsec$), the isovelocities on one side and the other of this axis display the expected  ``V'' shape associated with differential rotation.  The bar being aligned with the major axis, no asymmetrical distortions are expected in the isovelocities but a small stretching between the galaxy centre and the ``V'' shape is observed. As we go further out of the galaxy, the ``V'' shape of the isovelocities rotates slightly along the P.A. on both sides of the galaxy (at $\sim$6550 \ km \ s$^{-1}$ for the northern side and at $\sim$6950 \ km \ s$^{-1}$ for the southern side). This twist within the optical radius, an obvious sign of interaction, is especially seen on the northern side.

\subsection{Rotation curves}
\label{rot_curves}

In the case of non-merging systems (systems where the disc and the bulge of both galaxies are distinct and where the discs of both galaxies are separated by a distance larger than their optical diameter), velocity fields are usually rather smooth and symmetric, resulting in symmetric and low-scattered rotation curves (RCs) up to a certain radius \citep{2004A&A...415..451F}. With this assumption in mind, the RC of each galaxy was computed in order to obtain a symmetric curve and to minimize scatter on each side of the curve, leading to an optimal set of kinematical parameters.  Our methodology is described in \citet{2004A&A...415..451F}. In order to minimize the influence of non-circular velocity components that increase toward the minor axis of the RC, both RC were computed  considering points on the velocity field within an angular sector of $ 36 ^\circ $  on each side of the galaxy's position angle (P.A.).
 
\begin{figure}
\centering
\includegraphics[width=0.9\textwidth]{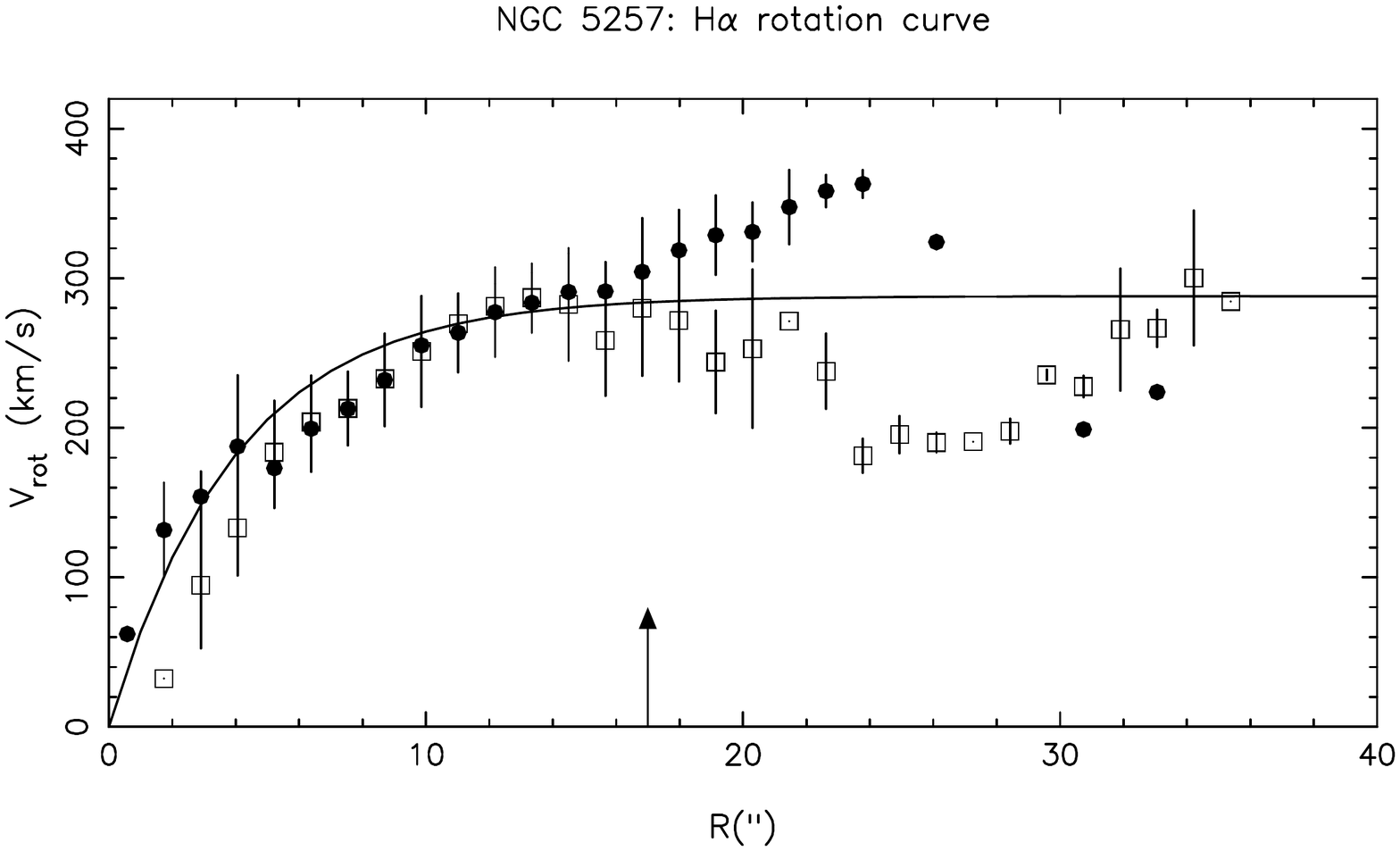}
\includegraphics[width=0.9\textwidth]{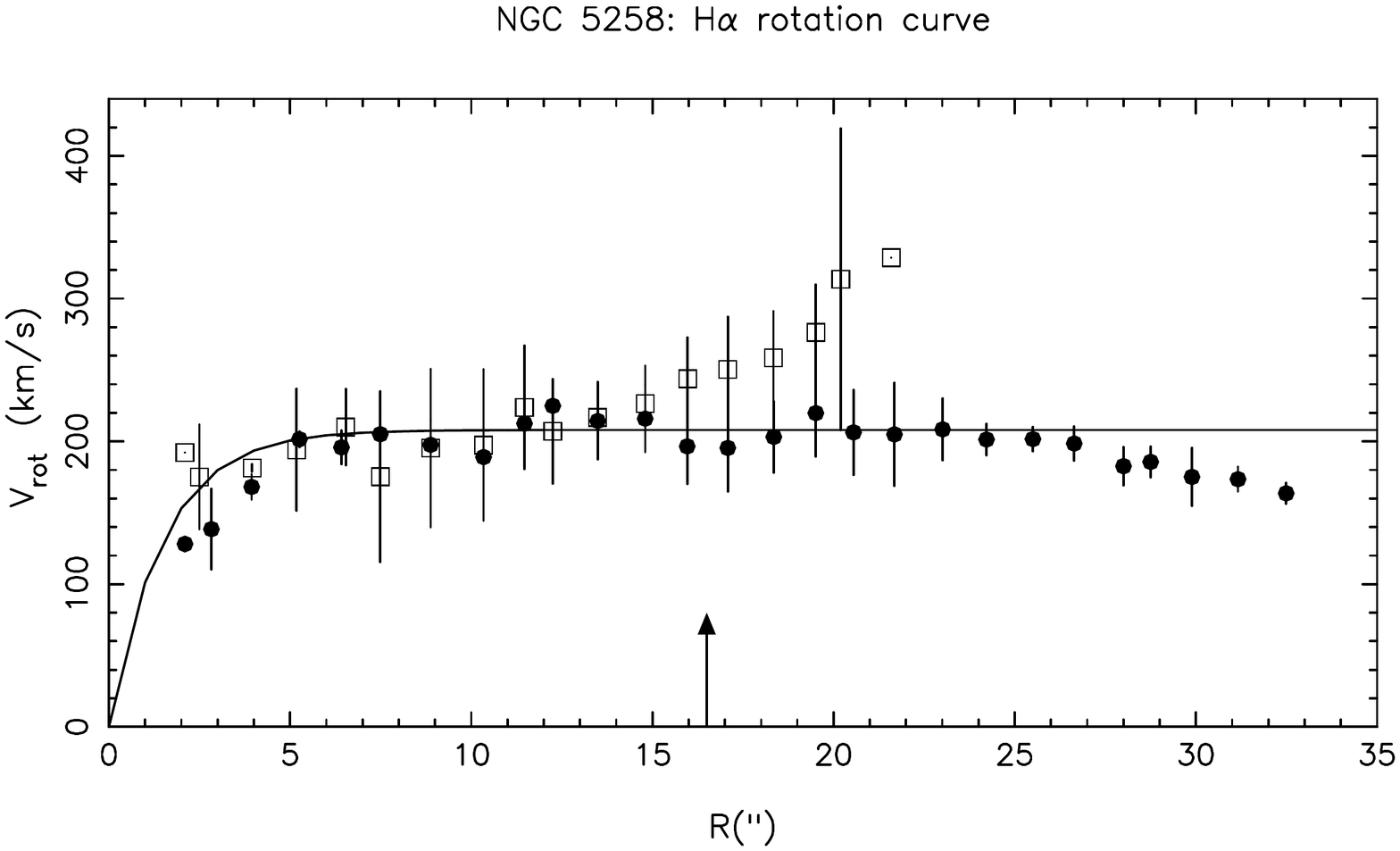}
\caption{Top: Rotation curve (RC) of NGC 5257 derived from scanning Fabry-Perot (FP) observations of the ionized gas. Dark circles represent velocities on the approaching side of the galaxy; light squares indicate velocities on the receding side of the galaxy. Bottom: Same as top panel for NGC 5258. 
  Solid lines show the parametric fit of each RC discussed in Sect. \ref{vit_res}.
  Arrows indicate the position of $R_{bif}$ discussed in Sect. \ref{rot_curves}.
}
\label{ngc5257-58_RCs}
\end{figure}

Figure \ref{ngc5257-58_RCs} shows the RCs of NGC 5257 and NGC 5258. The most symmetrical, smooth, and less-scattered RC, displaying a residual velocity field without systemic features (see Sect. \ref{vit_res}), was derived for both galaxies. The kinematical parameters used to derive the RC of each galaxy are presented in  Table \ref{glxs_kin+dyn-params}. Error bars give the dispersion of the rotation velocities computed for all the pixels found inside each elliptical ring having a width of about $\sim$ 1.5 $\arcsec$ and defined by the successive bins.  The presence of a bar induces non-circular motions in the central regions of the disc due to the streaming of the gas along the bar. If the bar is parallel to the major axis, as is the case for NGC 5258, the rotational velocities are underestimated in the central region of the galaxy (e.g. \citet{2008MNRAS.385..553D,2015MNRAS.454.3743R}). On the contrary, if the bar is perpendicular to the major axis, the measured rotational velocities are overestimated. For an intermediate case like NGC 5257, we do not expect any signature in the rotation curve. In our case, no correction has been applied to any of the RCs to take into account the effect of the bar, since this feature is fairly weak and small in both galaxies.

The following set of values was derived for NGC 5257: P.A. =  (95  $\pm$ 3)   $^\circ$,  $i$ =  (58 $\pm$ 5)   $^\circ$ , and  $V_{sys} =$ (6812 $ \pm$ 5) \  km \  s$^{-1} $. The kinematical centre of the galaxy lies $1.3\arcsec$ ($\sim630 $ pc) north of the  photometric centre of the HST image in the F814W filter. However, this difference is not significant considering the seeing of our FP observations. Except for the innermost regions (R $<$ 2.4  kpc, 5 $\arcsec$), the RC is symmetric up to a radius  $R = 17.0 \arcsec $ (8.2 kpc) where the rotation velocity reaches $V_{rot} = $ 290 \ km \ s$^{- 1}$. At this radius, which we shall call $R_{bif}$, the curve bifurcates. The rotation velocity on the approaching side of the galaxy decreases, while for the receding side, the velocity slowly increases.

The RC of NGC 5258 was derived using the following set of values: P.A. =  (218  $\pm$ 5)   $^\circ$,  $i$ =  (57 $\pm$ 4)  $^\circ$ , and  $V_{sys} =$ (6762 $ \pm $ 5) \  km \  s$^{-1} $. The kinematical centre of the galaxy lies $0.8\arcsec$ ($\sim390 $ pc) north-east of the  photometric centre of the HST image in the F814W filter. Again, this difference is not significant considering the seeing of our FP observations. The RC  is symmetric up to $R = 16 \arcsec$ (7.8 kpc), except for the points within the  inner $ 3.0 \arcsec$ (1.5 kpc). After that radius  where  $V_{rot} \sim$220 \  km \  s$^{-1} $,  the velocities on the receding side of the galaxy increase, while those on the approaching side remain fairly constant. We shall also call this radius $R_{bif}$. Table \ref{corrs-ref} summarizes the features of each galaxy discussed in Sects. \ref{Halpha}, \ref{VFs}, and \ref{rot_curves}.

\begin{table*}
\caption{Kinematical Parameters of NGC 5257 and NGC 5258 from this work.}
\centering 
\label{glxs_kin+dyn-params}
\begin{tabular}{l l l l l}
\hline
\hline
\noalign{\smallskip}
Parameter              & \ \ \ \  &    NGC 5257  & \ \ \ \ \ \ \ \ \ \ \ &  NGC 5258  \\
\noalign{\smallskip}
\hline
\noalign{\smallskip}

Heliocentric systemic velocity  ($km \  s^{-1}$)  & &  $6812 \pm 5$  & &  $6762 \pm 5$   \\

P.A. ($^\circ$)                                   & &  $95 \pm 3$    & &   $218 \pm 5$    \\

Kinematical inclination ($^\circ$)                & &  $58 \pm 5$    & &   $57 \pm 4$     \\

                                                 & &          & &       \\

 $R_{bif}$  in arcsec (kpc)            & &  $ 17 \ (8.23) $     & &  $ 16 \ (7.76)  $      \\

 $V_{bif}$ ($km \ s^{-1}$)                    & &  $290$             & &  $220$ \\

 $\Omega_p$ ($km \ s^{-1} \ kpc^{-1}$)        & &  $20 \pm 3$                     & &  ...   \\  

 $R_{cor}$ in arcsec (kpc)             & &  $ 15 \pm 2 \  (7.27 \pm 0.97)  $  & &  ...   \\ 

            & &          & &       \\

Maximum rotation velocity ($km \ s^{-1}$)  & &   $290$\tablefootmark{a}         & &   $210$\tablefootmark{a}         \\

                                          & &   $360 \pm 10$\tablefootmark{b}  & &    $330 \pm 5$\tablefootmark{b}   \\

                                          & &   $325$\tablefootmark{c}         & &    $320$\tablefootmark{c}         \\

\noalign{\smallskip} \hline
\end{tabular}
\tablefoot{
\tablefoottext{a} {From analytical fit of H$\alpha$ RC before bifurcation (Fig. \ref{ngc5257-58_RCs}).}
\tablefoottext{b} {Point with highest velocity in  H$\alpha$ RC considering all points (Fig. \ref{RC+morph}).}
\tablefoottext{c} {From fit of multi-wavelength RC (Fig. \ref{massmod}).}
}
\end{table*}

\begin{landscape}
\begin{table*}
\caption{H$\alpha$ emission, velocity fields, and rotation curve features for NGC 5257 and NGC 5258.}
\label{corrs-ref}
\centering 
\begin{tabular}{l c c c c c c c c c}
  \hline
    \hline
 \noalign{\smallskip}
    \multicolumn{10}{c}{H$\alpha$ IMAGE}\\  
 \noalign{\smallskip}

\hline
\noalign{\smallskip}
Galaxy             &     &   & Prominent HII regions                         &  &  Emission in central parts  &  & Emission along tidal tail &  & Emission along tidal bridge  \\
\noalign{\smallskip}
\hline
\noalign{\smallskip}
 NGC 5257           &    &   & YES -along both spiral arms                   &  &          NO                 &  &  YES                   &  & YES -at the beginning of the feature \\ 
\noalign{\smallskip}
 NGC 5258            &   &   & YES -only one                                 &  &          NO                 &  &   NO                   &  & NO \\  

 \noalign{\smallskip}
  \hline
    \hline
\noalign{\smallskip}
\noalign{\smallskip}

\multicolumn{10}{c}{VELOCITY FIELD}\\
\noalign{\smallskip}

\hline
\noalign{\smallskip} 
Galaxy         &       &   & Central isovelocities                         &  &  Presence of bar (P.A.)       &  &  Outer isovelocities   &  & Visual signs of interaction  \\
\noalign{\smallskip}
\hline
\noalign{\smallskip}
NGC 5257        &       &   & Parallel to each other,            &  &  YES   &  &  Follow rotating disc   &  &  YES -on tidal tail \\
                 &      &   & not perpendicular to major axis   &  &         &  &  pattern                &  &                     \\      
\noalign{\smallskip}
NGC 5258          &     &   & Parallel to each other,            &  &  YES   &  &  Isovelocities ``V'' shape      &  &  YES -along disc    \\
                   &    &   & perpendicular to major axis        &  &         &  & rotates along P.A.           &  &                    \\  
 \noalign{\smallskip}
\hline
\hline
\noalign{\smallskip}
\noalign{\smallskip}
  \multicolumn{10}{c}{H$\alpha$ ROTATION CURVE}\\  
\noalign{\smallskip} 
  \hline
\noalign{\smallskip} 
Galaxy           &      &   &   Kinematic and Photometric  &  &  Inner asymmetries   &  &  Symmetrical   &  &  Bifurcated   \\
                  &     &   &   centres match              &  &                      &  &                &  &               \\
\noalign{\smallskip}
\hline
\noalign{\smallskip}
NGC 5257           &    &   &  YES -within PUMA seeing     &  &  From 0\arcsec to 5\arcsec     &  & From 5\arcsec to 17\arcsec    &  &  From 17\arcsec \\
\noalign{\smallskip}
NGC 5258            &   &   &  YES -within PUMA seeing     &  &  From 0\arcsec to 2.5\arcsec   &  & From 2.5\arcsec to 16\arcsec  &  &  From 16\arcsec \\
\noalign{\smallskip}
\hline
\hline
\end{tabular}
\end{table*}

\end{landscape}

\subsection{Non-circular motions}

\subsubsection{Velocity dispersions}
\label{fwhm}

The gas velocity dispersion, $\sigma$, for each pixel was computed from the H$\alpha$ velocity profiles after correction from instrumental and thermal widths, $\sigma_{inst} = FSR / (\mathcal F \times 2\sqrt {2 \ln 2} ) =$ 16.4 \ km \ s$^{-1}$ and $\sigma_{th} =$ 9.1 \  km \ s$^{-1}$, respectively. The latter value was estimated by assuming an electronic temperature of T$_{e}$ = 10$^4$ K in the expression $\sigma_{th}$=(kT$_{e}$/m$_{H}$)$^{1/2}$. Assuming that all the profiles (instrumental, thermal, and turbulent) are described by Gaussians, the velocity dispersion was estimated using $ \sigma = (\sigma_{obs}^2 - \sigma_{inst}^2 - \sigma_{th}^2)^{1/2}$.

\begin{figure*}
\centering
\includegraphics[width=0.5\textwidth]{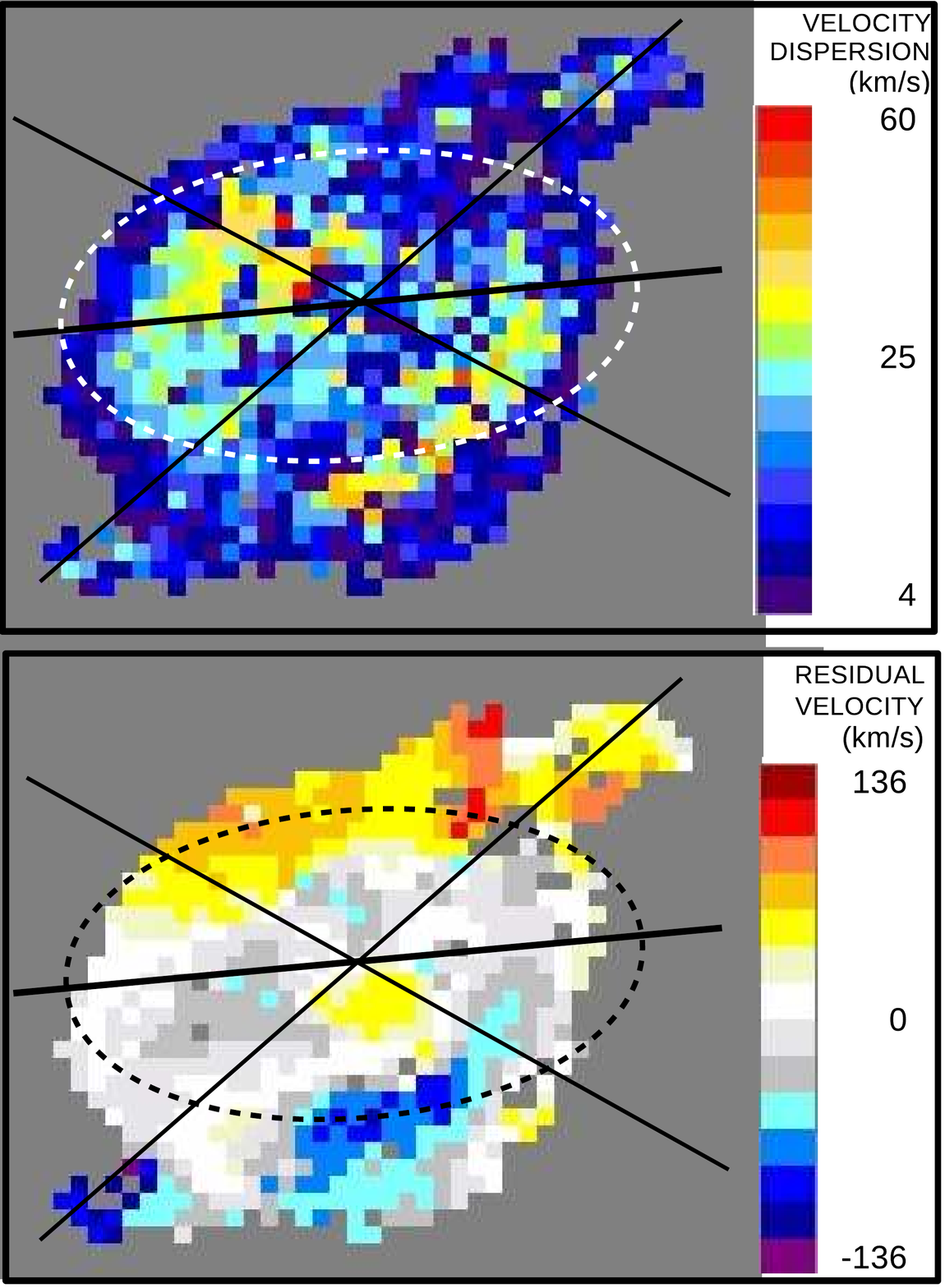}
\includegraphics[width=0.5\textwidth]{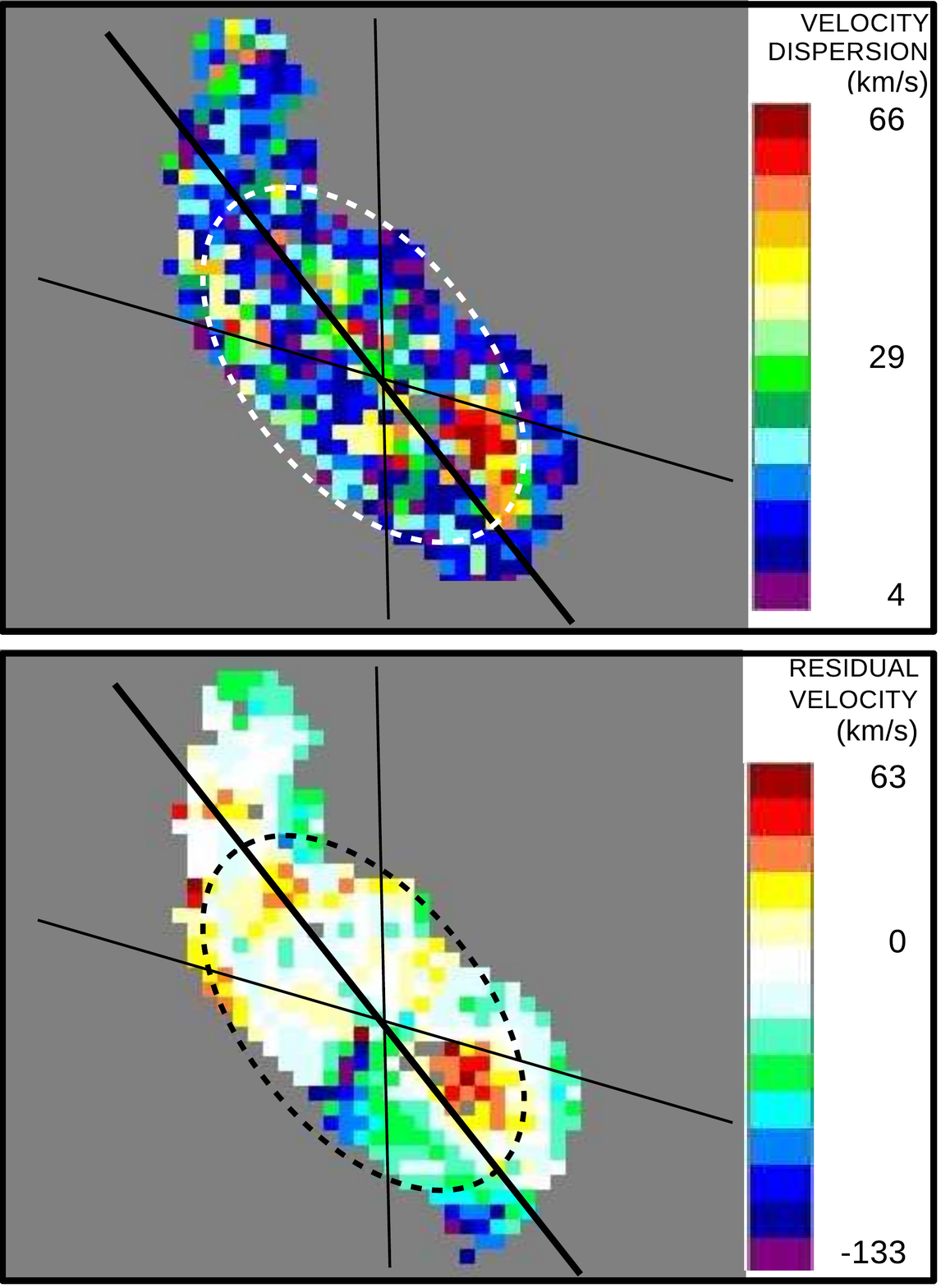}
\caption{Top left: Velocity dispersion map for NGC 5257. Bottom left: Residual velocities map of NGC 5257. Top right:  Velocity dispersion map for NGC 5258. Bottom right: Residual velocities map of NGC 5258.
  The bold dark line is the P.A. and the fainter black lines the angular sector used to compute the RC of each galaxy.
Ellipses indicate the location of the bifurcation radius, $R_{bif}$ , in the rotation curve of NGC 5257 and NGC 5258, respectively, discussed in Sect. \ref{rot_curves}.
}
\label{noncirc}
\end{figure*}

The top left panel of Fig.  \ref{noncirc} shows the velocity dispersion map of NGC 5257. High velocity dispersion values ranging between  30 \ km \ s$^{-1} $ and  55 \ km \ s$^{-1} $ are seen along both spiral arms.  The highest values match the location of the HII region in the eastern arm.  Velocity values larger than 35 \  km \ s$^{-1} $ are observed along the western spiral arm matching the location of the H$\alpha$ maxima.  Values of $\sim $ 25 \ km \ s$^{-1}$ are observed on the tip of this arm coinciding with the beginning of the tidal arm. The top right panel of Fig. \ref{noncirc} shows the velocity dispersion map of NGC 5258. Large values ($\sim $ 55 \ km \ s$^{-1}$) are seen on the south-western side of the galaxy matching the location of the large HII region. The highest velocity dispersions are asymmetrically observed on the southern part of the galaxy. 

\subsubsection{Residual velocity fields}
\label{vit_res}

The residual velocity field of each galaxy was obtained by subtracting an axisymmetric velocity field model from the observed  velocity field. The model  velocity field was derived by fitting the observed RC of the galaxy with a simple parametric function, $A \times (1-e^{-r/B})$, assuming it reflects pure circular motions.  A check has been done on this image to ensure that its pattern does not contain any features due to a bad determination of the systemic velocity, kinematic centre, P.A., and inclination as shown by \citet{1973MNRAS.163..163W}.  The bottom left panel of Fig.  \ref{noncirc} shows the residual velocities map for NGC 5257. A contribution of positive residual velocities ($\sim$40  km \ s$^{-1}$) is seen in the south-central part of the galaxy; no symmetric counterpart is observed on the other side of the major axis of the galaxy. Values ranging from 40 \ km \ s$^{-1}$ to 110 \ km \ s$^{-1}$ are seen on the northern arm of the galaxy associated with the presence of the large HII regions. Positive residual velocities are seen on the tip of this arm.  Negative values ranging between -40 \ km \ s$^{-1}$ and -100 \ km \ s$^{-1}$ follow the distribution of strong H$\alpha$ emission along the southern arm. Negative values of $\sim$-100 \ km \ s$^{-1}$ are also observed on the tip of this arm to the south-east.
For NGC 5258 (bottom right panel of Fig. \ref{noncirc}), positive residual velocities ranging between 30 \ km \ s$^{-1}$ and  63 \ km \ s$^{-1}$ are found at the location of the large HII region on the south side of the galaxy. Negative residual velocities ranging from -30 \ km \ s$^{-1}$ to -100 \ km \ s$^{-1}$ are located on the southern tip of the galaxy and negative values of $\sim$-60 \  km \ s$^{-1}$ are also detected on the northern arm of the object.

\subsection{Kinematic versus morphological features}
\label{kin_morph}

\begin{figure*}
\centering
\includegraphics[width=0.4\textwidth]{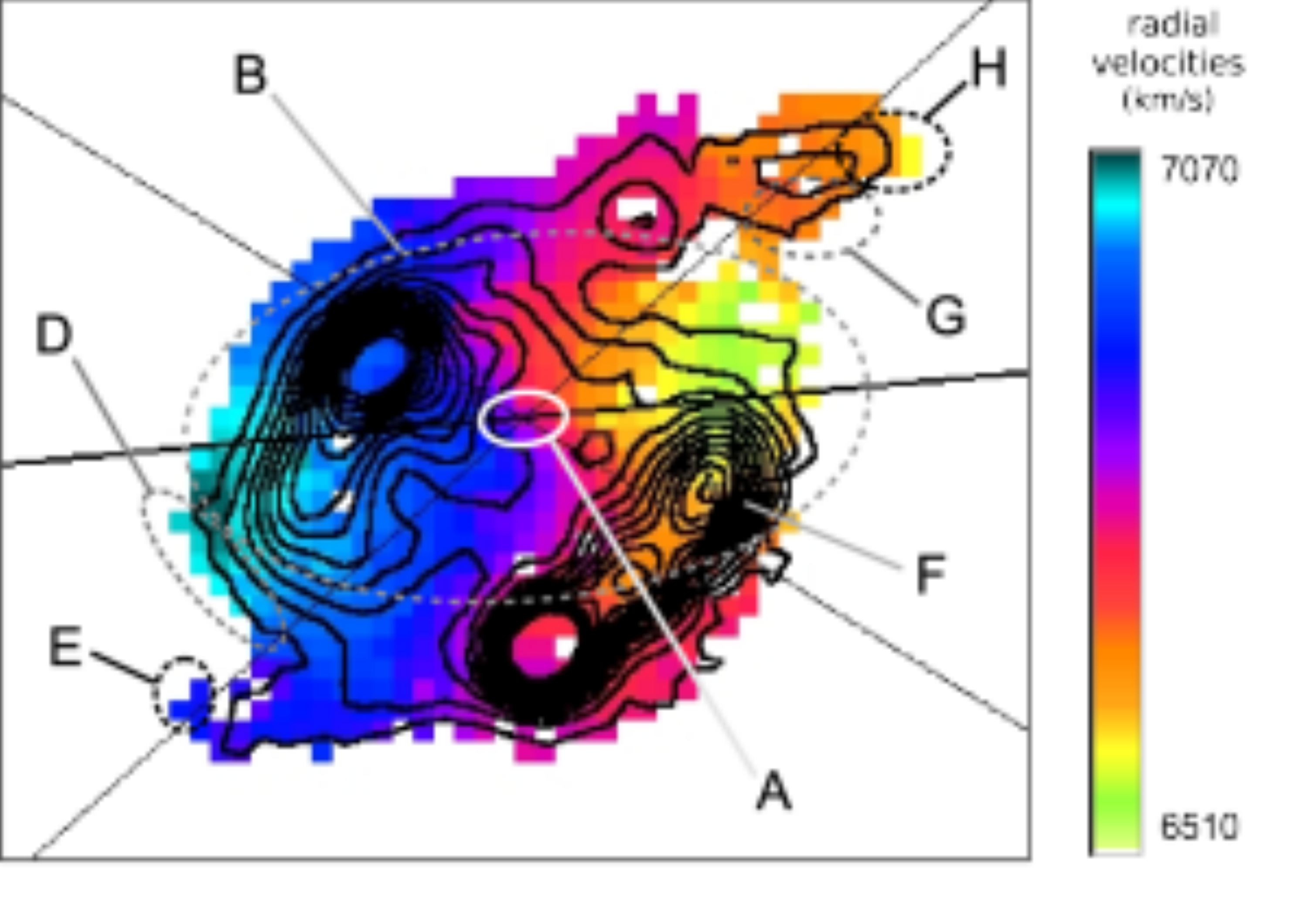}
\includegraphics[width=0.4\textwidth]{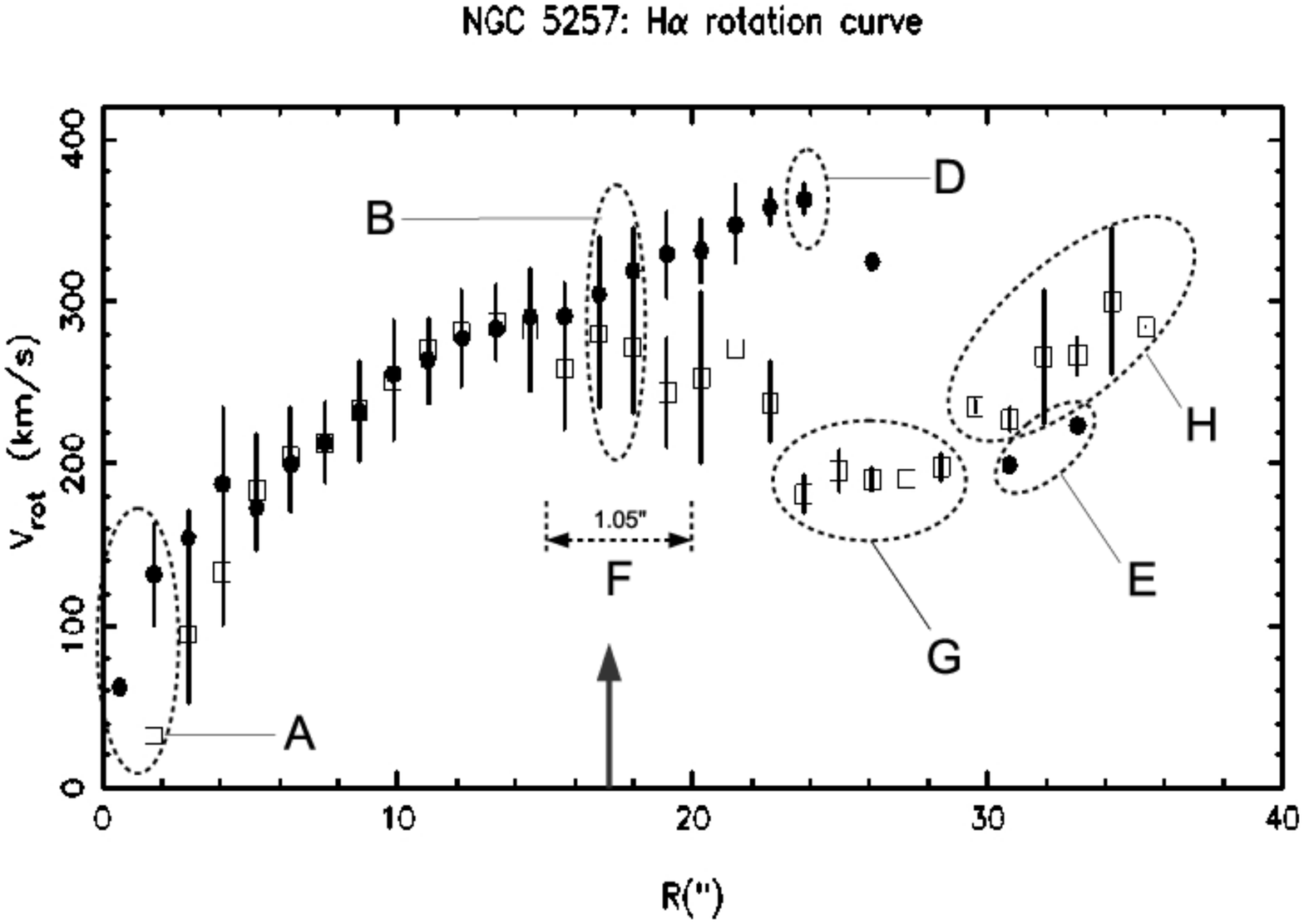}
\includegraphics[width=0.4\textwidth]{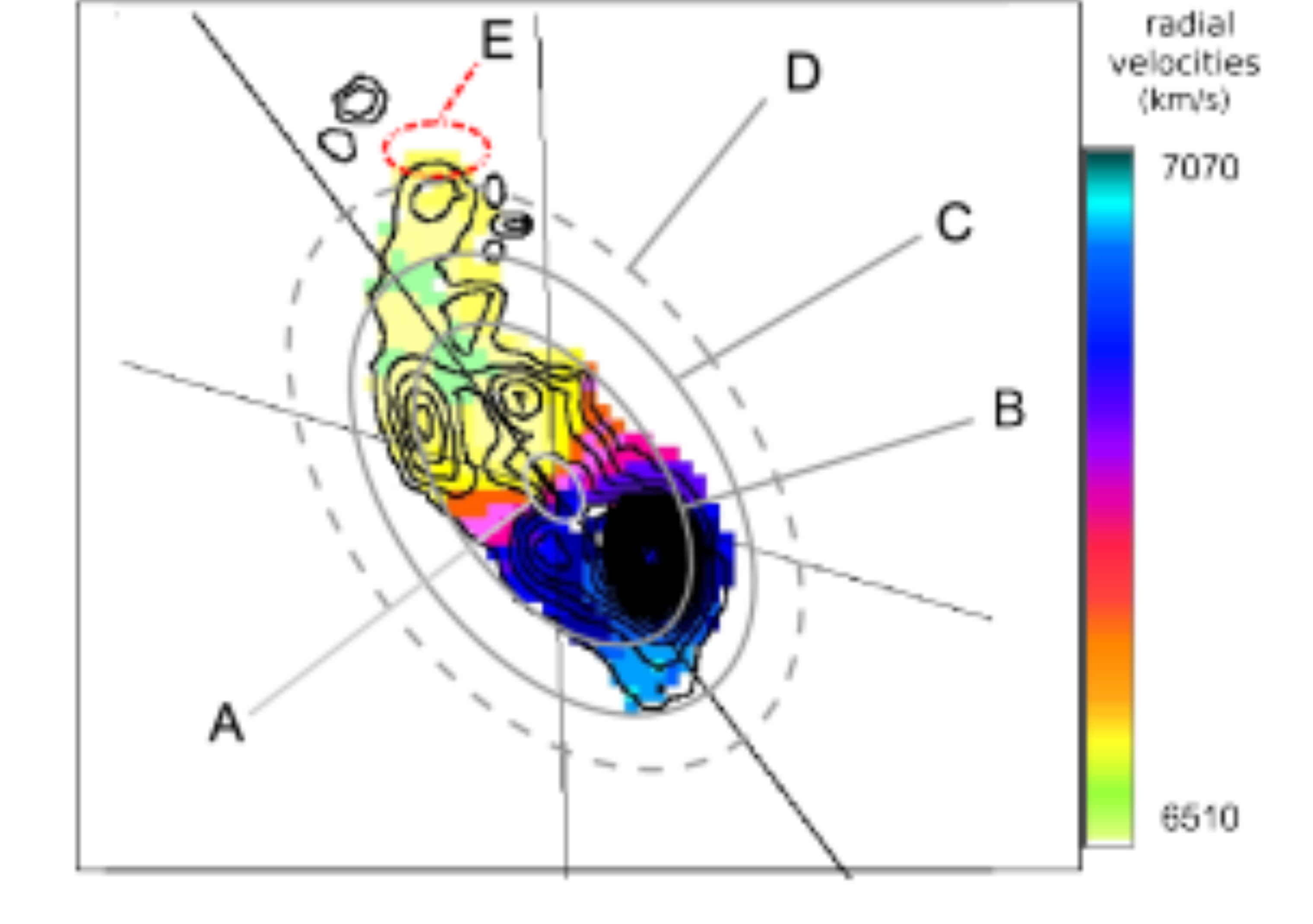}
\includegraphics[width=0.4\textwidth]{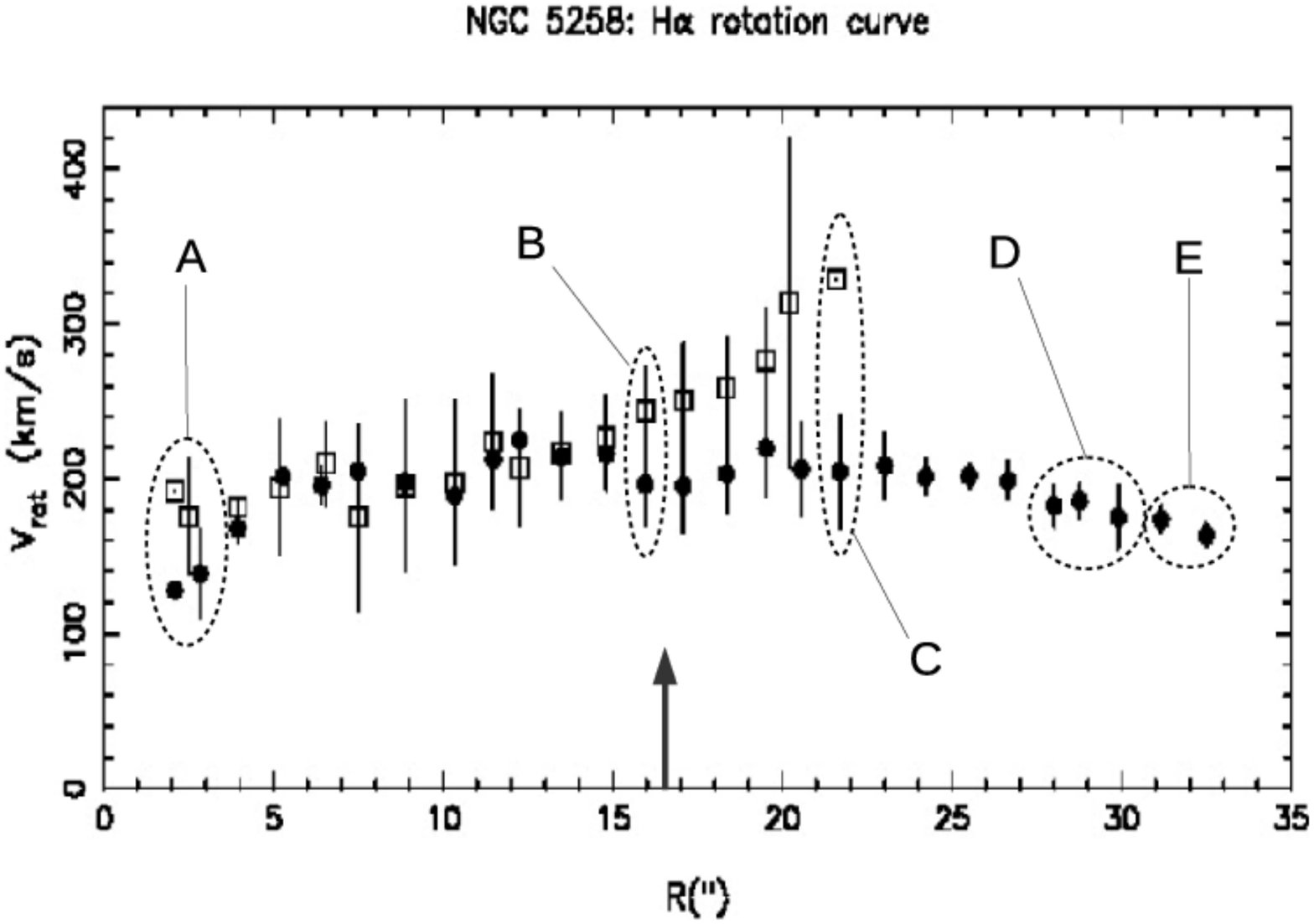}
\caption{Top left: Velocity field of NGC 5257 with monochromatic isophotes. Letters indicate features associated with variations in the rotation curve of the galaxy. Solid line indicates the galaxy's position angle (P.A.), the  slash-dotted lines indicate the angular sector from both sides of the major axis considered for the computation of the galaxy's RC. Top right: Rotation curve (RC) of NGC 5257 derived from scanning Fabry-Perot (FP) observations of the ionized gas. Dark circles represent velocities on the approaching side of the galaxy; light squares indicate velocities on the receding side of the galaxy. The regions and letters shown are associated with features in the rotation curve of the galaxy.  These are discussed in Sect. \ref{kin_morph}.
Bottom left: Same as top left panel for NGC 5258.
Bottom right: Same as top right panel for NGC 5258.
} 
\label{RC+morph}
\end{figure*}

Two-dimensional velocity fields of disc galaxies portray the motion of ionized gas all over the galaxy enabling us to match these motions with different morphological structures. One can determine to what extent the gas follows a circular motion around the centre of the galaxy or if there are important contributions from non-circular velocities in the radial, azimuthal, and vertical directions  due to the presence of these structures or to external perturbations. For this purpose,  we compared the monochromatic  H$\alpha$ image of each galaxy with its velocity map and its RC (Fig. \ref{RC+morph}).

For NGC 5257, the inner points of the RC show an asymmetry between the approaching and receding sides. This corresponds to part of the central bar (points A in the bottom panel of Fig. \ref{RC+morph} and region A in top panel of the same figure). From these points onwards, both sides of the RC remain symmetric up to a $R=17 \arcsec $  (8.2 kpc) reaching a rotational velocity of 290  \ km \ s$^{- 1}$ (point B in the RC). This radius matches the location of $R_{bif}$ defined in Sect. \ref{rot_curves}. For the approaching side the rotation velocity increases to 340 \ km \ s$^{- 1}$ at $R=24 \arcsec$ (point D in the RC). This radius corresponds to  the last points of the main disc of the galaxy on the receding side (region D in  the bottom panel of Fig. \ref{RC+morph}). Finally, $V_{rot}$ drops to a value of $\sim$200  \ km \ s$^{- 1}$ at a radius of  $ \sim$32$\arcsec $ (15.5 kpc)  (point E in the RC). This corresponds to the tip of the extended western arm of the galaxy in H$\alpha$ (region E). For the receding side of the RC (open squares in the bottom panel of Fig. \ref{RC+morph}), the velocity decreases as we approach the western edge of the main disc of the galaxy at  $17\arcsec $ (8.2 kpc) (ellipse B). In particular, this change seems to be associated with the strong  HII region marked with F in the top panel of Fig. \ref{RC+morph}. The velocity $V_{rot}$ reaches a low-velocity plateau that seems to be associated with parts of the eastern tidal arm (region G). Then it increases again from a radius of  $30\arcsec $ (14.5 kpc) and seems to be associated with  the tip of this tidal arm (region H).

For NGC 5258, the asymmetry between the receding and approaching sides in the central parts corresponds to region A in Fig. \ref{RC+morph}.  From a radius of $ 4\arcsec $ (1.9 kpc) to $ 16\arcsec $ (7.8 kpc) (point B in the RC - bottom panel of  Fig. \ref{RC+morph}), both sides are symmetric reaching a velocity of  220  \ km \ s$^{- 1}$.   From this radius onwards, the velocity values increase for the  receding side (open squares in the bottom panel of Fig. \ref{RC+morph}), while they remain fairly constant for the approaching side (dark circles in the same figure). This radius, $R_{bif}$, matches the edge of the main disc of the galaxy (ellipse B in the top panel of  Fig. \ref{RC+morph}).  For the receding side (open squares in bottom panel of Fig. \ref{RC+morph}), the velocity increases to 330  \ km \ s$^{- 1}$ at a radius of  $12\arcsec $ (5.8 kpc)  (point C in the RC). This corresponds to the southern tip of the galaxy where tidal effects along the arms of the galaxy begin to be seen (ellipse C). For the approaching side  (dark circles in the bottom panel of Fig. \ref{RC+morph}), the velocity remains  constant ($\sim$ 220  \ km \ s$^{- 1}$) up to $R=23 \arcsec $ (11.2 kpc), slightly decreasing from this radius to $27 \arcsec $ (13.1 kpc) (before region D in the bottom panel of Fig. \ref{RC+morph}). The velocity decreases along region D (top panel of Fig.  \ref{RC+morph}), reaching  a minimum value of 160  \ km \ s$^{- 1}$ at  $32\arcsec $ (15.5 kpc) (point E).

An ellipse with a major axis equal to  $R_{bif}$ was plotted on the velocity dispersion and residual velocity maps of each galaxy. For NGC 5257, large values of velocity dispersion fall inside the ellipse for the eastern spiral arm; for the western arm, two of the HII regions fall inside the ellipse, while the brightest one falls outside the ellipse. Important values of the residual velocity along the spiral arms follow the ellipse, except for the values at the tips of the arms. For NGC 5258, both large dispersion velocity and residual velocity values probably associated with the large HII region fall inside the ellipse.

\begin{figure*}
\centering
\includegraphics[width=0.8\textwidth]{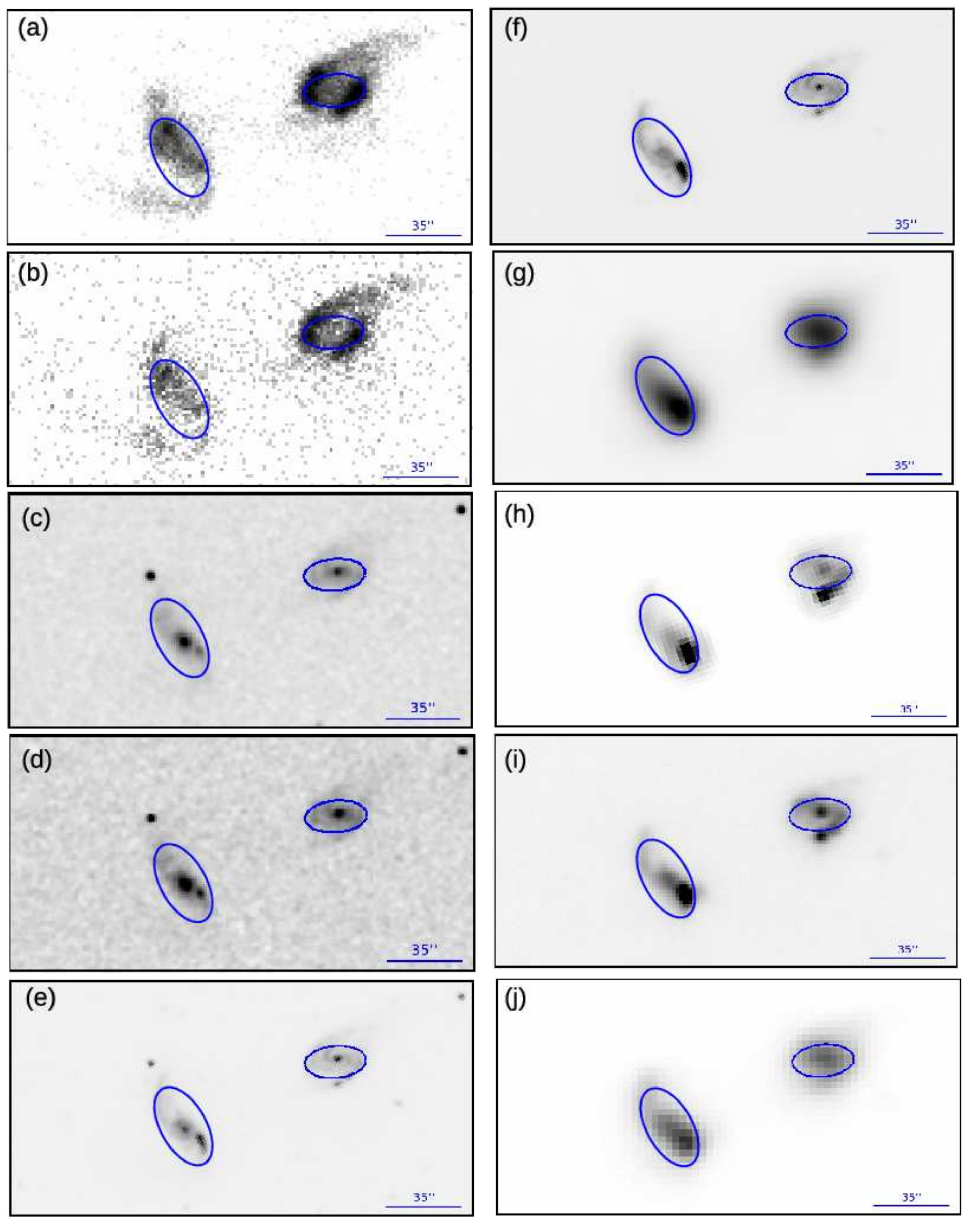}
\caption{ 
Near-ultraviolet (a) and far-ultraviolet (b) images of Arp 240 taken from $GALEX$ archive.
The $J$ image (c) and $K$ image (d) from $2MASS$ \citep{2006AJ....131.1163S}.   
The 4.5 $\mu$m (e) and  8 $\mu$m (f) images taken  with $Spitzer$-$IPAC$ camera \citep{2007AJ....133..791S};   
12 $\mu$m  (g) image taken with $WISE$-$PAC$ (\citet{2010AJ....140.1868W}; $IRSA$ catalogue); 
24 $\mu$m (h) image taken with $Spitzer$-$MIPS$ \citep{2007AJ....133..791S};    
60 $\mu$m (i) and 160 $\mu$m (j) images taken with $Herschel$ "blue'' channel and "red'' channel, respectively (Proposal ID KPGT esturm 1).
Ellipses indicate the location of the bifurcation radius, $R_{bif}$, in the rotation curve of NGC 5257, discussed in Sects. \ref{rot_curves}, \ref{kin_morph}, and \ref{omega_p}. 
North is to the top, east is to the left.
} 
\label{arp240_bands}
\end{figure*}

For both galaxies, we also plotted an ellipse with a major axis equal to  $R_{bif}$ on images of the galaxy at different wavelengths (Fig. \ref{arp240_bands}). The P.A. and inclination values of each ellipse were those used to derive the RC in Sect. \ref{rot_curves}. The position of the ellipse follows the location of $R_{bif}$ for each galaxy. For NGC 5257, the main disc of the galaxy is confined to the ellipse in the $K$-band, 4.5, 8.0, 12, and 160 $\mu$m images. For the optical images, only the southern part of the disc is limited by the ellipse. The maximum near ultraviolet (NUV) emission lies outside the ellipse, as well as the MIR and FIR maxima. For NGC 5258, the ellipse encompasses the ultraviolet (UV) emission from the main disc of the galaxy, while the tidally stretched arms lie outside the ellipse. For the NIR images, the main disc lies inside the ellipse. For the northern side of the galaxy, the ellipse matches the point where the northern arm bifurcates into the western and northern tidal extensions. The brightest HII regions on the southern side of the galaxy lie inside the ellipse.

\subsection{Pattern speed determination}
\label{omega_p}

The pattern speed, $\Omega_p$, is the speed of a density perturbation that moves along at a speed different from the speed of the objects within it. This density perturbation can be due to the presence of a bar, a spiral density wave, and/or the presence of a companion. The value of $\Omega_p$ can be estimated from kinematic measurements without adopting any specific dynamical model: the Tremaine-Weinberg method (TW; \citet{1984ApJ...282L...5T})   requires only knowledge of the distribution of intensity and velocity of a component that reacts to the density wave
and that obeys the continuity equation. Though it has been mainly used with the stellar component of disc galaxies, the applicability of the TW method to H$\alpha$ emission has been shown in \citet{2005ApJ...632..253H}, \citet{2007ApJ...667L.137F,2009ApJ...704.1657F},   and \cite{2009ApJ...702..392G}  for isolated galaxies.  We assumed this method is still  valid for slightly perturbed, non-merging interacting galaxies for which the ionized gas mostly traces the global kinematics of each galaxy, such as occurs in  Arp 240.
For this pair of galaxies,  both the disc and the bulge are distinguishable and the velocity field of each galaxy resembles that of a rotating disc. They do not appear to be strongly perturbed and no common HI envelope is observed. Although tidal features are seen, they do not seem to have important effects in the inner parts of each galaxy. The validity of this assumption will be discussed in Sect. \ref{disc}. 

Using H$\alpha$ emission as tracer, $\Omega_p$ was derived from the  relation \citep{1995MNRAS.274..933M} 

$$ \Omega_p = { 1 \over {sin \ i}} \ \  { \langle V(x) \rangle \over { \langle x \rangle} } \eqno,(1) $$

where $\langle V(x) \rangle $ is the intensity-weighed, line-of-sight average velocity and $\langle x \rangle $ the intensity-weighted average position of the tracer along an ``aperture'' that is parallel to the apparent major axis of the galaxy. The pattern speed value $\Omega_p$ is obtained by fitting a straight line to these points. The slope of this line equals $\Omega_p \times \sin i$. We set the origin of the Cartesian coordinate system at the derived kinematical centre of each galaxy  and extracted the line-of-sight velocity along successive slits parallel to the major axis in two-pixel intervals. The derived velocities within each simulated slit were ``collapsed'' into one single averaged value,  $\langle V(x) \rangle$.

\begin{figure*}
\centering            
\includegraphics[width=0.8\textwidth]{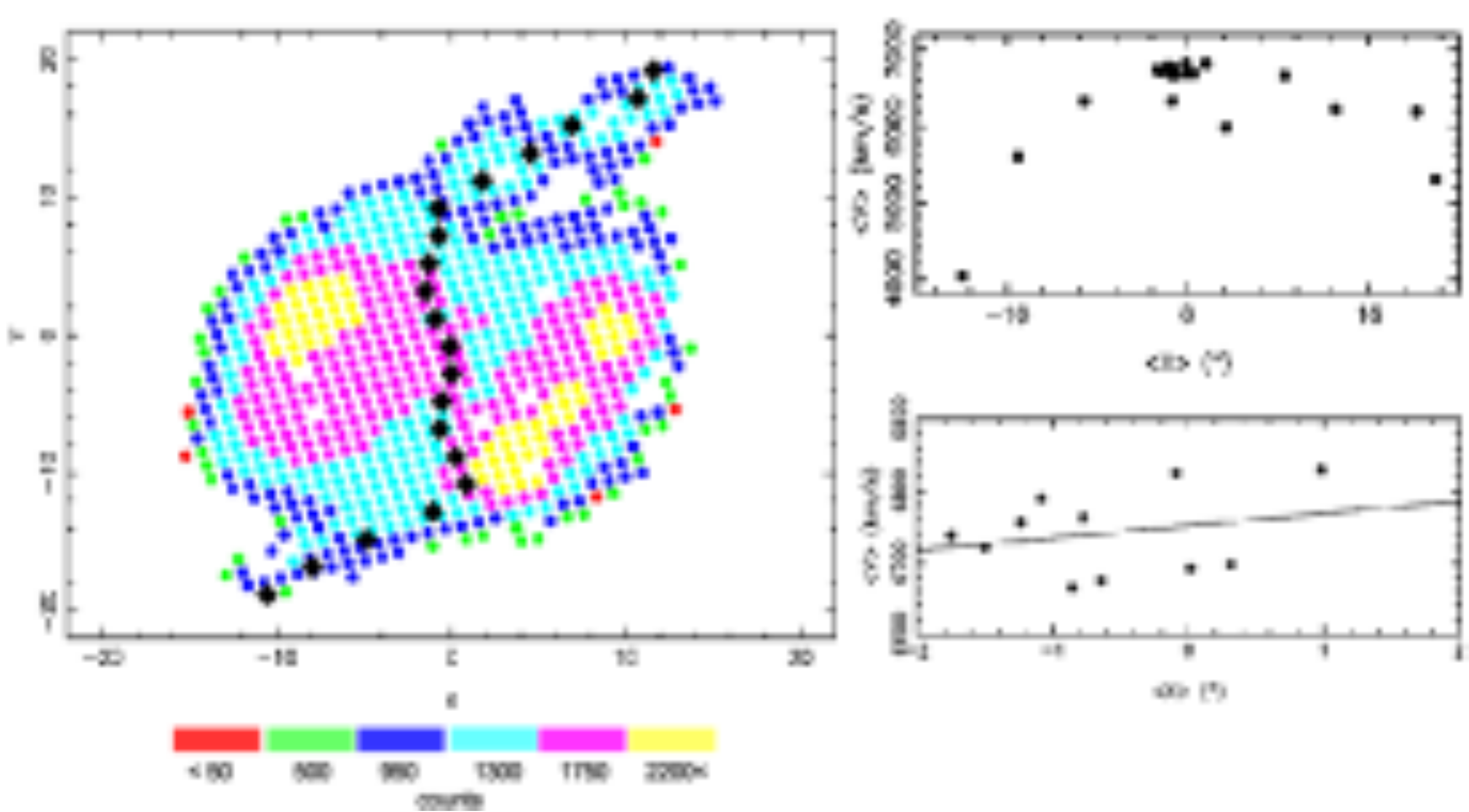}
\caption{Left panel: Mean Y-position versus mean X-position for the determination of the pattern speed of NGC 5257 using the Tremaine-Weinberg (TW) method. 
The galaxy has been rotated so that the P.A. lies along the X axis.
Dotted colours trace the monochromatic (H$\alpha$) emission of the galaxy.
Top right panel: $\langle V(x) \rangle$ versus $\langle x  \rangle$ for NGC 5257.
Bottom right panel: Zoom of the $\langle V(x) \rangle$ versus $\langle x  \rangle$ plot for NGC 5257.
The variables $\langle V(x) \rangle$ and $\langle x  \rangle$ are described in Sect. \ref{omega_p}.}
\label{ngc5257_omega-p}
\end{figure*}

For NGC 5257, twenty averaged points were derived.  The intensity-weighted average position per slit is shown in the left panel of Fig. \ref{ngc5257_omega-p}. The five upper averaged points in the plot and the four lower averaged points were derived using considerably less points along the slit than for the averaged points on the main disc of the galaxy, plus they seem to be associated with the beginning of the tidal bridge and the tidal tail of the galaxy. For this reason, they were not considered for the computation of  $\Omega_p$. Also, since we are interested in deriving $\Omega_p$ for a density perturbation (if any) on the main disc of the galaxy, we did not consider the two innermost averaged points since they are most likely associated with the kinematics of the bar. Once this pre-selection was done, a line was fitted to these points using linear least-squares (bottom right panel of Fig. \ref{ngc5257_omega-p}). The slope of this line equals $\Omega_p \times \sin i$. Considering the inclination value given in Sect. \ref{rot_curves}, we find $\Omega_p$ = 20 $\pm$ 3 \ km \ s$^{-1}$ \ arcsec$^{-1}$. The error computation was done considering the error on the inclination, on the mean velocity, on the mean position along the line of nodes, and on the P.A. For NGC 5258, the H$\alpha$ emission is mostly confined to the strong SF event on the southern arm of the galaxy,  therefore it cannot be considered a good kinematical tracer and the TW method cannot be used to estimate $\Omega_p$.

\begin{figure}
\centering
\includegraphics[width=0.8\textwidth]{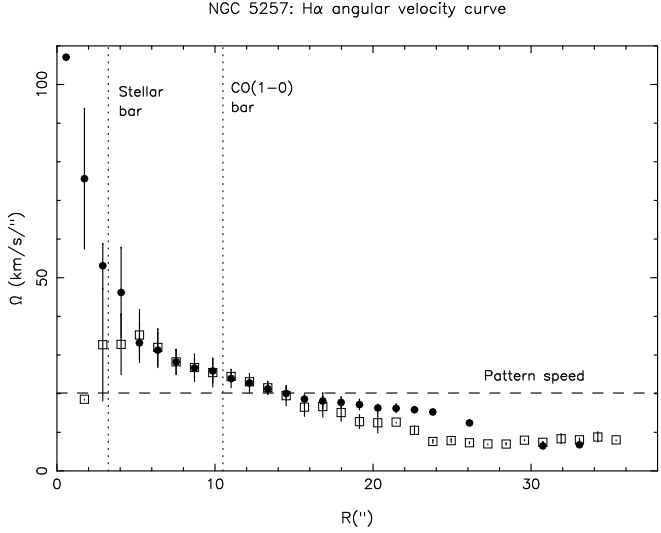}
\caption{Angular velocity for NGC 5257. Indicated on the figure are the location of both the stellar and CO(1-0) bar, as well as the pattern speed of the disc of the galaxy derived in Sect.  \ref{omega_p}.
}
\label{omegas}
\end{figure}

The value of the derived pattern speed for NGC 5257 was compared to the motion of the ionized gas in that galaxy.  Figure \ref{omegas} shows the H$\alpha$ angular velocity curve for each side of NGC 5257.  Horizontal dashed lines indicate the value of the derived pattern speed.  The radius at which the pattern speed and the RC intersect gives the position of the corotation radius of the galaxy, R$_{cor}$. This velocity matches the velocity of the ionized gas where the RC bifurcates at (15 $\pm$ 2) \arcsec, while the position of the stellar bar matches the point where the RC becomes symmetric.

\section{Dynamical results}
\label{dyns}

\subsection{Multi-wavelength rotation curve}
\label{multi-L_RC}

\begin{figure}
\centering
\includegraphics[width=0.8\textwidth]{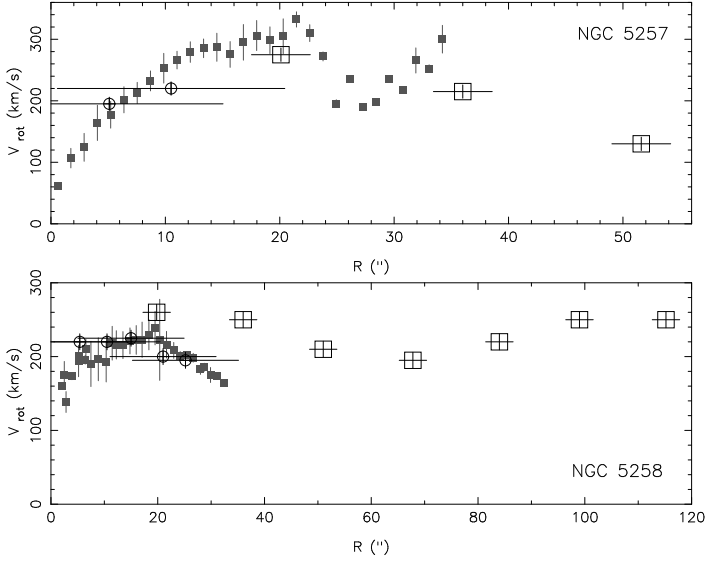}
\caption{Multi-wavelength rotation curves of NGC 5257 (top panel) and NGC 5258 (bottom panel) displaying all observed points. Small squares in the inner parts of the curve  correspond to optical Fabry-Perot H$\alpha$ observations. Empty circles  in the inner parts correspond to the CO(1-0) curve derived by \citet{2005ApJS..158....1I}. Empty squares in the  outer parts correspond to the HI curve also derived by \citet{2005ApJS..158....1I}. }
  \label{arp240_multiL-all}
\end{figure}

Along with the H$\alpha$ kinematic information from our FP observations, we used CO(1-0) and HI observations by \citet{2005ApJS..158....1I} in order to build a multi-wavelength rotation curve for each galaxy.  In order to compare the different curves for both galaxies, we plotted the rotation velocities of \citet{2005ApJS..158....1I} RCs using the inclination of the optical RC  (see Sect. \ref{rot_curves}). Figure \ref{arp240_multiL-all} shows the resulting multi-wavelength rotation curves for both galaxies.
For NGC 5257, the RCs superpose within the error bars up to $R=20 \arcsec $ (9.7 kpc) with $V_{rot} \sim$ 300 \ km \ s$^{-1}$. However, at larger radii, the optical RC decreases abruptly to $V_{rot} \sim$ 200 \ km \ s$^{-1}$ from $24 \arcsec $ (11.6 kpc) to $30 \arcsec $ (14.5 kpc), then it increases to 300 \ km \ s$^{-1}$ at  $R=35 \arcsec $ (17.0 kpc). Due to the beam size of the observations, no HI points are observed between $20 \arcsec $ (9.7 kpc) and $36 \arcsec $ (17.5 kpc). The HI RC declines in velocity ($ \Delta V>$100 \ km \ s$^{-1}$) at large radii ($>31.5\arcsec=$15.3 \ kpc ). According to \citet{2005ApJS..158....1I}, this could be due to a significant warp in the outer parts of the HI disc, but we think that the decreasing RC is more likely due to streaming motions related to the interaction rather than circular motions out of the plane of the disc. For NGC 5258, the optical and CO(1-0) RCs superpose smoothly within the error bars.  Both the optical and  CO(1-0) RCs begin to decrease at  $R=20 \arcsec $ (9.7 kpc); at $R=40 \arcsec $ (19.4 kpc) the HI curve shows a large oscillation, from  260 \ km \ s$^{-1}$  to  200 \ km \ s$^{-1}$ and back, in a $60 \arcsec $ (29.1  kpc) interval. After $R=100 \arcsec $ (48.5 kpc), the HI curve remains constant at 250 \ km \ s$^{-1}$. 

\subsection{Mass estimates through rotation curves}
\label{mass-est-RCs}

Making use of the possibility to disentangle circular from non-circular velocities in the optical RC, the multi-wavelength RCs were ``cleaned'' of points associated with non-circular motions (see Sect. \ref{kin_morph}) and points associated with features in the outer parts of the HI discs (Sect. \ref{multi-L_RC}).  For NGC 5257, we removed the outer points of the averaged H$\alpha$ RC lying after the bifurcation radius, $R_{bif}$.  For the HI points, we removed the last two points of the RC of \citet{2005ApJS..158....1I}, which correspond to the decreasing part of the RC.  For NGC 5258, we removed all points in the optical curve lying outside a radius of  $\sim20\arcsec =$ 9.7 \ kpc where the averaged H$\alpha$ RC starts to decrease. We also removed the CO($1 - 0$) points  lying beyond that radius since they showed an abrupt decrease in velocity. For the HI points, we removed points beyond  $40\arcsec $ (19.4 kpc) since they  correspond to the beginning of the extended  HI tidal tail detected in the eastern side of the galaxy (Fig. 3 in \citet{2005ApJS..158....1I}). Using these composite ``clean'' RCs, a range of possible masses was computed for each galaxy using the method by \citet{1983A&A...125..394L} according to which the total mass of a galaxy within a radius $R$ lies between $ 0.6$ (the case of a disc-like mass distribution) and $1.0$ $ \times \ (RV_{max}^2(R) / G) $ (the case of a spheroidal mass distribution). For NGC 5257, we considered the rotation velocity at the HI point before the velocity decreases (V = 270 \ km \ s$^{-1}$) to estimate the mass within $R = 20\arcsec $ = 9.7 kpc = 0.37 \ $ R_{25} $.  The range of masses within this radius equals $0.9$ to $1.6 \times 10^{11} M_\odot$. For NGC 5258, a rotation velocity of  250 \ km \ s$^{-1}$ was considered  to estimate the mass within  the last HI point before the extended tidal tail, $R = 38 \arcsec = $ 18.4 kpc = 0.72 \ $ R_{25}$. The range of masses within this radius is $1.6$ to $2.7 \times 10^{11} M_\odot$. For the sake of comparison, we computed the mass ratio of the galaxies within the same radius in $D_{25} / 2$ units.  The mass ratio, $M_{NGC5257} / M_{NGC5258}$, within a radius of $0.37 \ R_{25}$ goes from $0.5$ to $1.6$. On average,  $M_{NGC5257} / M_{NGC5258} \sim 0.9 $ within a radius of $0.37 \ R_{25}$. These results are shown in Table \ref{mass-estimates}.

\begin{table*}
\caption{Mass estimates for NGC 5257 and NGC 5258.}
  \centering 
\label{mass-estimates}
\begin{tabular}{l l l l l}
\hline
\hline
\noalign{\smallskip}
 Inside R=0.37 D$_{25}$/2             & \ \ \ \  &    NGC 5257  & \ \ \ \ \ \ \ \ \ \ \ &   NGC 5258   \\
\noalign{\smallskip}
\hline
\noalign{\smallskip}

R=0.37 D$_{25}$/2    in arcsec (kpc)         & &   $20.0 \ (9.70)$               & &  $  18.4 \ (8.94) $        \\

V at 0.37 D$_{25}$/2 ($ km \ s^{-1}$)    & &   $270$              & &  $ 290$         \\

Dynamical mass  ($ 10^{11} \ M_\odot$)   & &   $0.9 - 1.6$\tablefootmark{a}   & &  $1.5 - 2.5$\tablefootmark{a} \\

                                                       & &  $ 1.48$\tablefootmark{b}       & &  $1.02$\tablefootmark{b} \\

 $ M_{disc}$\tablefootmark{b}  ($10^{11} \ M_\odot$)     & &  $ 0.358$  & &  $0.238$ \\

 $ M_{halo}$\tablefootmark{b}   ($10^{11} \ M_\odot$)    & &  $ 1.12$  & &  $ 0.779$ \\

Mass ratio (M$_{NGC5257}$  / M$_{NGC5258}$)    & &        & $0.5 - 1.6$\tablefootmark{a}  &     \\

                                                                         & &        & $1.45$\tablefootmark{b}        &     \\
\hline
\hline
\noalign{\smallskip}
Inside furthest HI point on RC         & \ \ \ \  & NGC 5257  & \ \ \ \ \ \ \ \ \ \ \ &  NGC 5258  \\
before velocity decrease, R$_{HI}$     & \ \ \ \  &           & \ \ \ \ \ \ \ \ \ \ \ &            \\

 \noalign{\smallskip}
\hline
\noalign{\smallskip}

R$_{max}$   in arcsec             & &   $20.0$ (0.38 $D_{25}/2$)         & &      $38.0$ (0.72 $D_{25}/2$)        \\

V at R$_{HI}$ ($km \ s^{-1}$)     & &   $ 270$                           & &   $ 250$         \\

Dynamical mass  ($10^{11} \ M_\odot$)          & &  $ 0.9 - 1.6$\tablefootmark{a}  & &  $ 1.6 - 2.7$\tablefootmark{a} \\

                                             & &  $1.48$\tablefootmark{b}       & &  $2.87$\tablefootmark{b} \\

 $M_{disc}$\tablefootmark{b}  ($10^{11} \ M_\odot$)              & &  $0.358$  & &  $0.406$ \\

 $M_{halo}$\tablefootmark{b}  ($10^{11} \ M_\odot$)              & &  $1.12$   & &  $2.46$  \\

\noalign{\smallskip} \hline
\end{tabular}
\tablefoot{
\tablefoottext{a} {Using \citet{1983A&A...125..394L} method.}
\tablefoottext{b} {From RC decomposition.} 
}

\end{table*}

\subsection{Mass distribution and dark matter halo profile}
\label{mass_mod}

\begin{figure}
\centering
\includegraphics[width=0.8\textwidth]{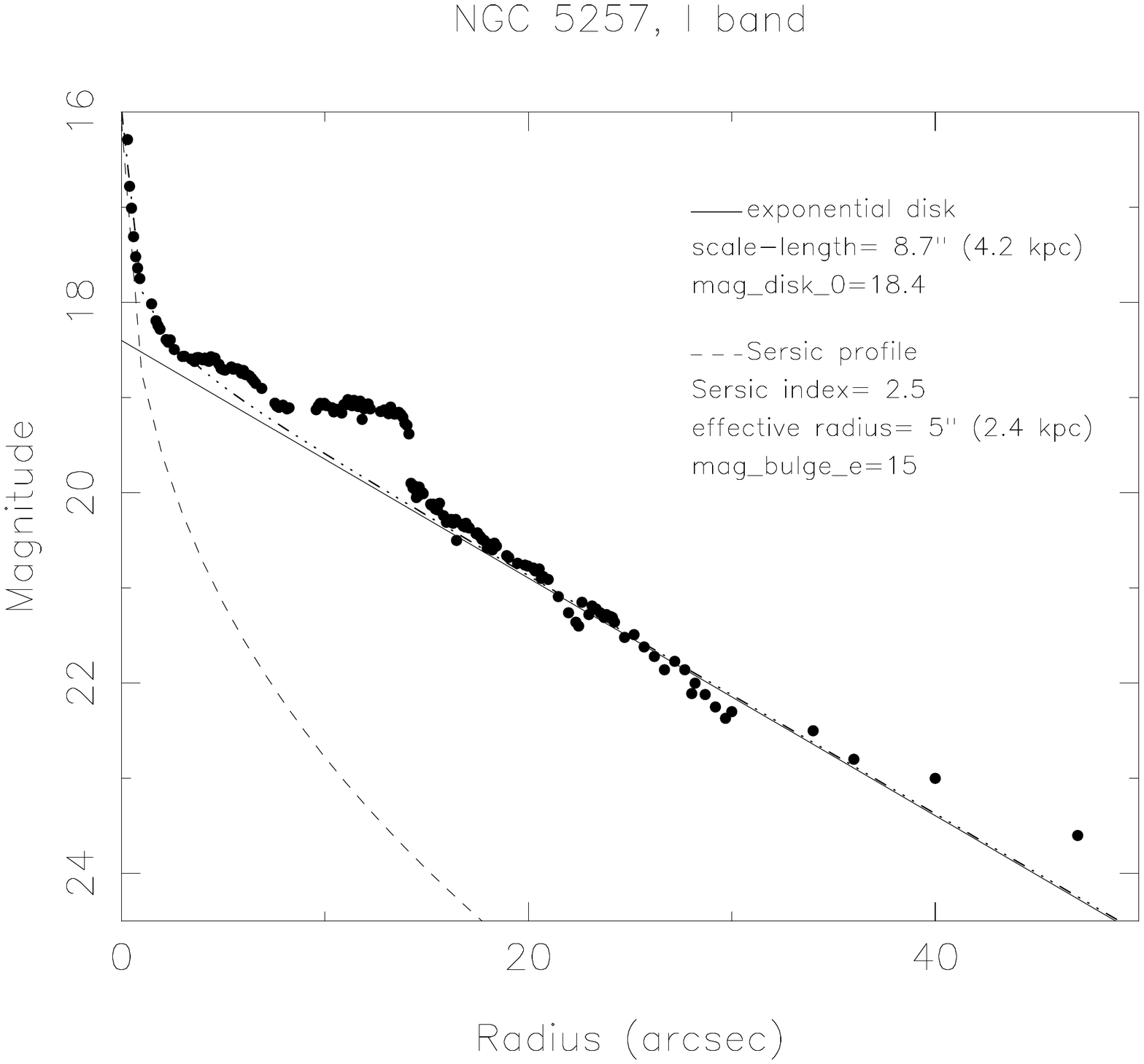}
\includegraphics[width=0.8\textwidth]{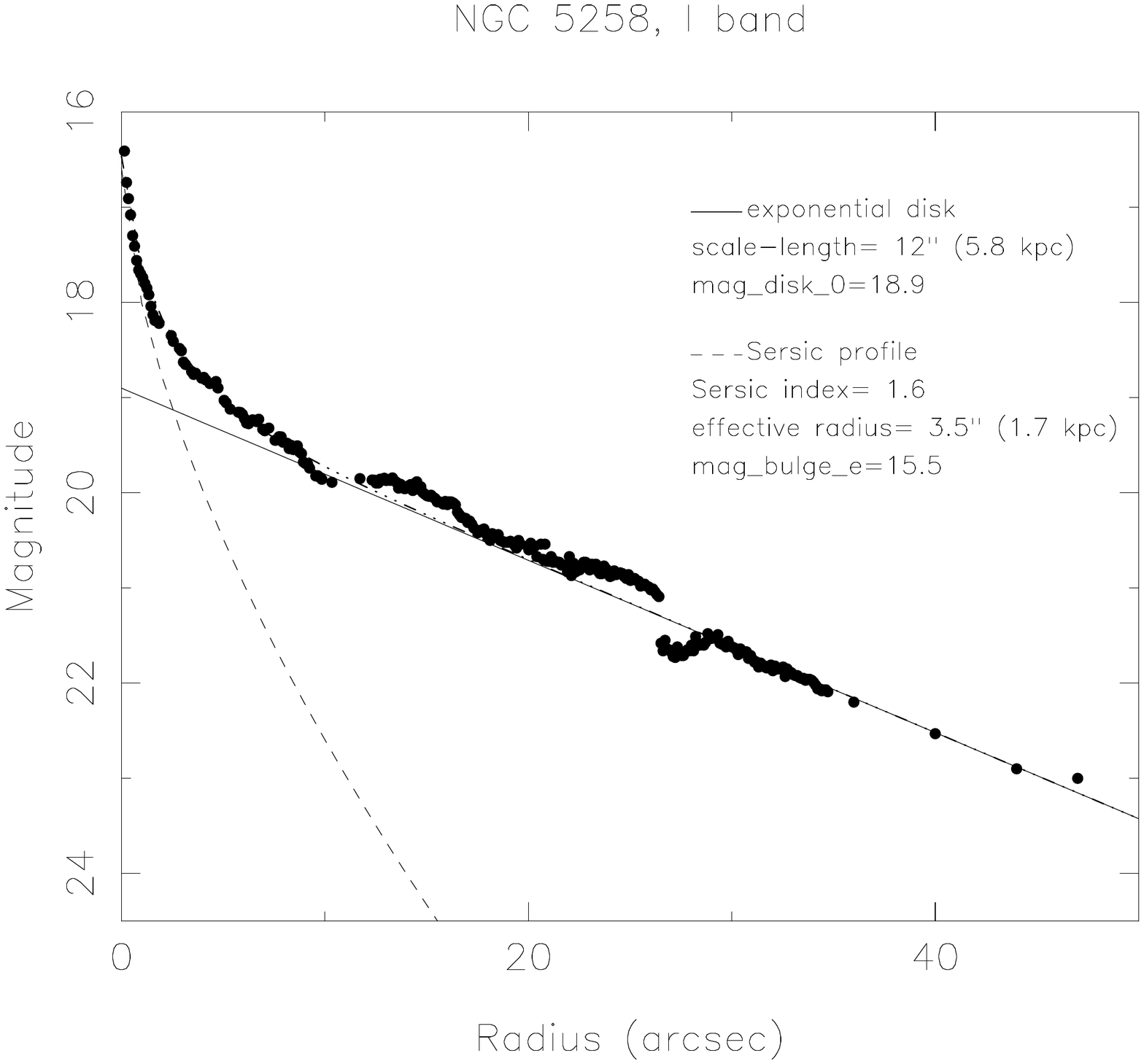}
\caption{Left panel: Surface brightness profile of NGC 5257 in the HST-ACS F814W filter.
Image taken from the HST archive, proposal 10592 by Aaron Evans.
Right panel: Surface brightness profile of NGC 5258 in the HST-ACS F814W filter.
Profiles have been decomposed in an exponential disc and a Sersic bulge component.
The dashed-dotted line shows the fit of each profile.
} 
\label{SBs}
\end{figure}

The mass distribution in each galaxy was estimated using the mass model described in \citet{2001AJ....121.1952B}, which considers both the light distribution of the galaxy (stars and HI) and a theoretical dark halo profile to compute a RC that best fits the observed one. The mass-to-light ratio of the disc, $(M/L)_{disc}$, and of the bulge, $(M/L)_{bulge}$, the dark matter (DM) halo central density, $\rho_{0}$, and core radius, $R_{0}$, are free parameters.  In order to determine the light distribution from each galaxy, we derived the surface brightness (SB) profile of each galaxy from the {\it F814W} HST images using the STSDAS package ISOPHOTE in IRAF.\footnote{IRAF is distributed by the National Optical Astronomy Observatories, operated by the Association of Universities for Research in Astronomy, Inc., under cooperative agreement with the National Science Foundation.} The light distribution was fitted with an exponential profile and a Sersic profile \citep{1968adga.book.....S} in order to determine the contribution of the bulge and that of the stellar disc. Both profiles show the presence of the large HII regions in both galaxies. For NGC 5257, the northern HII region is responsible for the surface brightness increase around 12\arcsec.   The HI surface density distribution was taken from \citet{2005ApJS..158....1I}. Figure \ref{SBs} shows the SB profile of NGC 5257 (top) and of NGC 5258 (bottom).

Two types of DM halos were considered: a pseudo-isothermal sphere (pISO; \citet{1987PhDT.......199B}) and a Navarro-Frenk-White (NFW) profile \citep{1996ApJ...462..563N}. Both a maximal and a non-maximal disc were considered.\footnote{We considered the ``maximal disc'' definition of \citet{1997ApJ...483..103S} for which  $ 85 \% \pm 10\%$ of the total rotational support of the galaxy at a radius $ 2.2 \times scale \ radius$ is contributed by the stellar disc mass component.}  The model was fitted to the "cleaned'' multi-wavelength RC\footnote{For both galaxies, the innermost points of the CO(1-0) RC were removed due to their large position error bars.} of each galaxy described in Sect. \ref{mass-est-RCs}. Table \ref{massmodpar} shows the mass model parameters for each fit. 

In our case, all halo models (with or without maximal disc) give similar $\chi^2$ values. However, $M/L$ values vary from one fit to the other -see Table \ref{mass-mods}. We used these values to choose a ''best'' fit considering the multi-wavelength images of each galaxy discussed in Section \ref{kin_morph}.   For NGC 5257, the best fit was achieved considering $(M/L)_I$ values of 3.5 for both the disc and the bulge. For this fit, a pISO halo and a maximal disc (with an important contribution of the bulge) were considered. For the other choices of DM halo and maximal or non-maximal disc,  $(M/L)_I$ values are either low for both the disc and the bulge components or extremely high for the disc and extremely low for the bulge (see Table \ref{mass-mods}). 

For NGC 5258, the best fit was achieved considering $(M/L)_I$ values of 3.2 for the bulge and 3.1 for the disc. In this case, a pISO halo and a maximal disc were considered. A pISO halo and maximal disc option result in low values for $(M/L)_I$. An NFW halo and a maximal disc give large values for $(M/L)_I$; while an NFW halo with a non-maximal disc result in low  $(M/L)_I$ values for both the bulge and the disc. Values are shown in Table \ref{mass-mods}.

\begin{figure}
\includegraphics[width=0.5\textwidth]{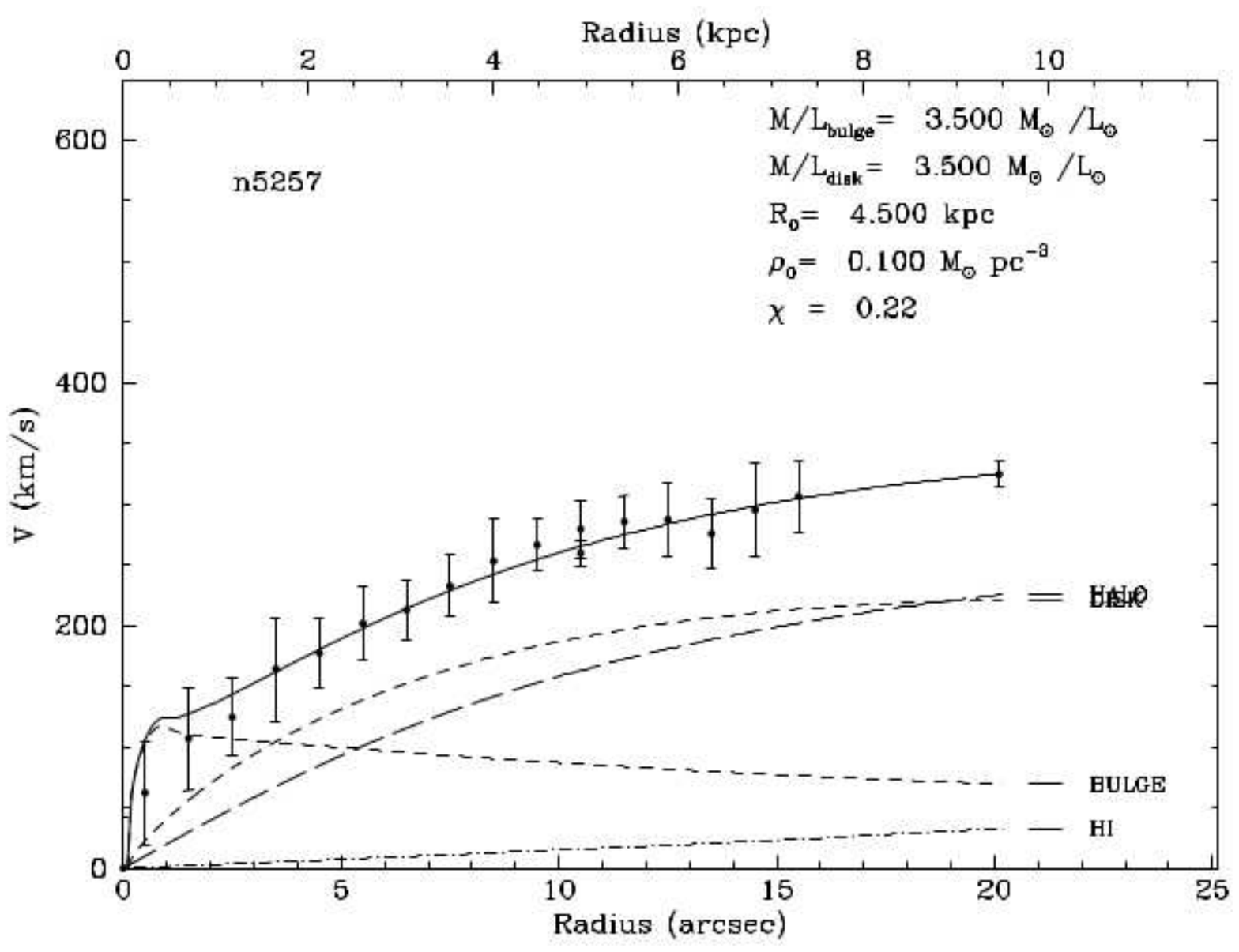}
\includegraphics[width=0.5\textwidth]{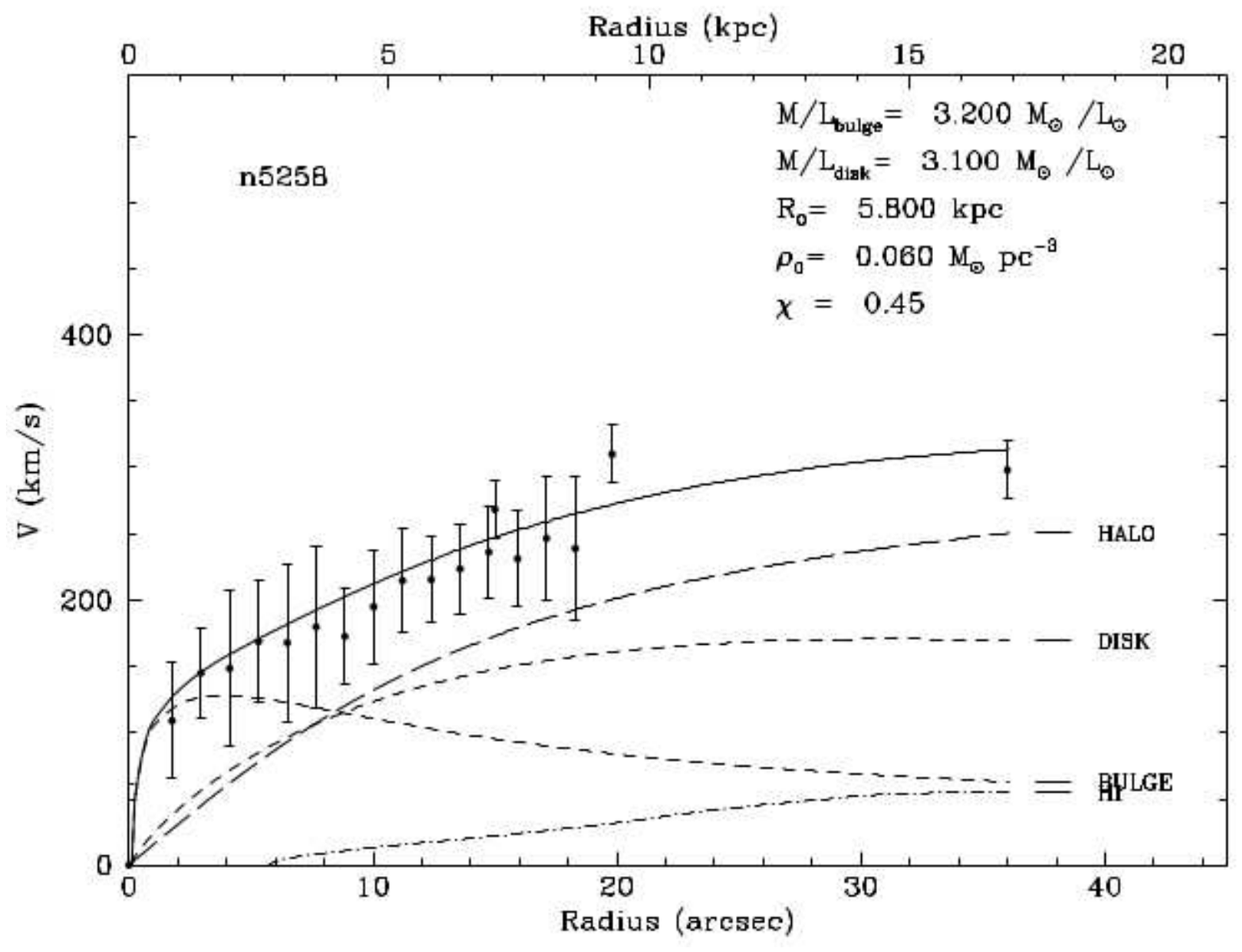}
\caption{Left: Best mass model fit for the multi-wavelength rotation curve for NGC 5257. Right: Best mass model fit for the multi-wavelength rotation curve for NGC 5258.} 
\label{massmod}
\end{figure}

\begin{landscape}

\begin{flushleft}
\begin{table*}
\centering \caption{Mass models parameters for NGC 5257 and NGC 5258 from
fits of the multi-wavelength rotation curve considering only points associated with circular motions.} \label{massmodpar}
\label{mass-mods}
\begin{tabular} {c c c c c c c c c c c c c c c c}
\hline\hline
 \noalign{\smallskip}
Galaxy & & Type of & & Maximal & & $(M/L)_{bulge}$ & & $(M/L)_{disc}$ & & $R_0$ & & $ \rho_0$ & & $\chi^2$ \\
       & & halo   & &  disc   & & (I-band) & & (I-band) & & (kpc) & & ($10^{-3} M_\odot/pc^{-3}$) & &  \\
\hline
NGC 5257 & & pISO & & yes & & 3.50  & & 3.50 & & 4.5   & & 0.01  & & 0.22 \\
         & & pISO & & no  & & 1.75  & & 0.50 & & 2.6   & & 0.40  & & 0.12 \\
         & & NFW  & & yes & & 0.01  & & 6.30 & & 9.0   & & 0.05  & & 0.12 \\
         & & NFW  & & no  & & 0.50  & & 0.50 & & 16.8  & & 0.04  & & 0.19 \\

NGC 5258 & & pISO & & yes & & 3.20   & & 3.10  & & 5.8   & & 0.06  & & 0.45 \\
         & & pISO & & no  & & 2.40   & & 1.55  & & 4.5   & & 0.11  & & 0.43 \\
         & & NFW  & & yes & & 4.30   & & 3.70  & & 45.0  & & 0.71  & & 0.70 \\
         & & NFW  & & no  & & 1.00   & & 1.00  & & 33.6  & & 0.01  & & 0.50 \\

\hline
\end{tabular}
\end{table*}
\end{flushleft}

\end{landscape}

\section{Spectral energy distribution}
\label{sed}

\begin{figure}
\centering
\includegraphics[width=0.65\textwidth]{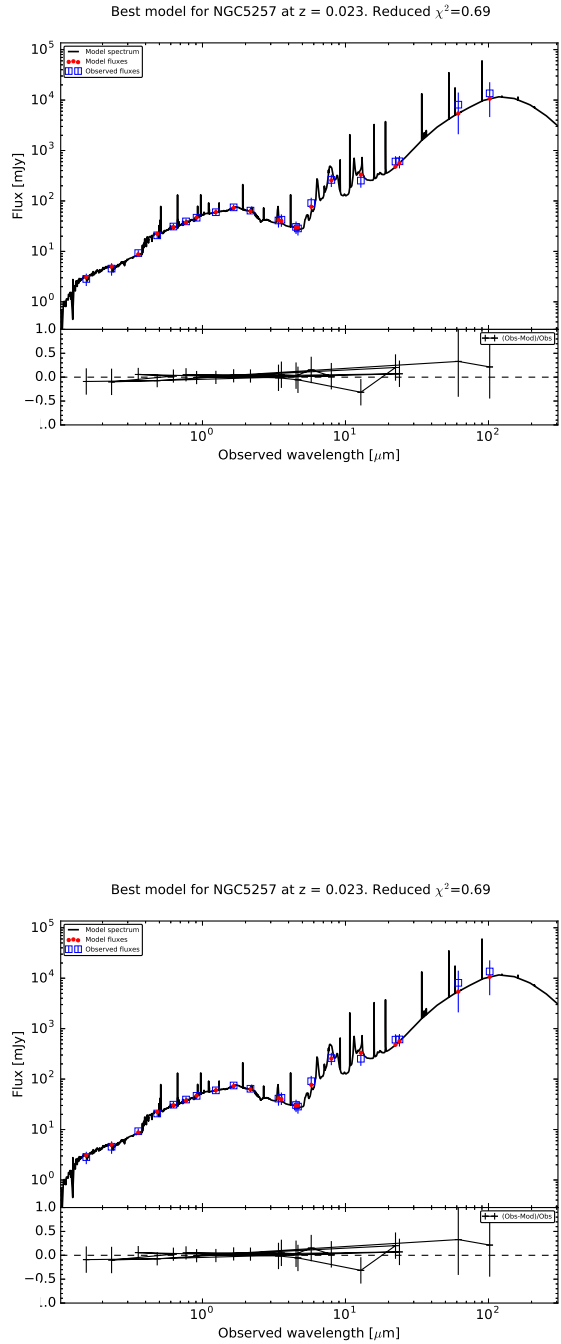}
\includegraphics[width=0.65\textwidth]{NGC5257_best_model.pdf}
\caption{Top left panel: Spectral energy distribution (SED) for NGC 5257.  Top right panel: SED for NGC 5258. A SED compiled using stellar synthesis model CIGALE \citep{2009A&A...507.1793N}  has been fitted to each SED.
Empty blue squares correspond to observed fluxes, red dots indicate model fluxes, and the black solid line corresponds to the model spectrum.
Bottom panels: Normalized error fit versus wavelength for each galaxy.
}
\label{arp240_SEDs}
\end{figure}

Figure \ref{arp240_SEDs} shows the spectral energy distribution (SED) of NGC 5257 (top panel) and of NGC 5258 (bottom panel) using the photometry values from \citet{2014ApJS..212...18B}.
FUV (far ultraviolet) and NUV fluxes are taken from the Galaxy Evolution Explorer (GALEX), u-, g-, r-, i-, z-band fluxes from the Sloan Digital Sky Survey (SDSS), J-, H-, Ks-band fluxes from the Two Micron All-Sky Survey (2MASS), along with Spitzer Space Telescope Infrared Array Camera (SST-IRAC) 3.6, 4.5, 5.8, 8 micron fluxes, Wide-field Infrared Survey Explorer (WISE) 3.6, 4.6, 12, 22 micron fluxes, and IRAS 60 and 100 micron fluxes.

In order to fit the observed SEDs, we used the SED-fitting code CIGALE \citep{2009A&A...507.1793N}.  CIGALE creates synthetic spectra from the \citet{2003MNRAS.344.1000B} stellar population models  followed by the attenuation of the stellar population spectra using a synthetic Calzetti-based attenuation law \citep{2000ApJ...533..682C} before adding the dust emission as given by the IR SED library built from the \citet{2002ApJ...576..159D} templates, which is particularly useful for fitting the SED of LIRGs. The code uses two single stellar populations with an \citet{1955ApJ...121..161S} initial mass function (IMF) to reproduce an old and a young stellar population, since two stellar components have been found necessary to reconstruct more accurate SFRs in actively star-forming galaxies  \citep[e.g.][]{2006ApJ...646..107E,2009ApJS..184..100L}. The model assumes an SFH with exponentially decreasing SFR: $\tau_1$ and $\tau_2$ represent the $e$-folding time of the old and young stellar populations, respectively (see \citet{2011A&A...525A.150G} for a detailed description). Initial parameters for the fit of the SED of each galaxy are shown in Table \ref{SEDparam}.
The base consists of two single stellar populations (SSP) from \citet{2003MNRAS.344.1000B} and a $\tau$-exponentially declining star formation history with $\tau$ = 10.0, 20.0, 50.0, 100.0, 200.0, 400.0, 600.0, 800.0,1000.0 Myr for the young population  and  $\tau$ =1, 2, 3, 4, 5, 6, 7 Gyr for the old population. For both galaxies, an age of 13 Gyr was given as a fixed value for the old stellar population, while for the young stellar population, the initial age values were  10.0, 20.0, 50.0, 100.0, 200.0, 400.0, 600.0, 800.0, and 1000.0 Myr. Following \citet{2009A&A...507.1793N}, a solar metallicity Z = 0.02 Z$_\odot$ was selected. Initially the young versus old stellar population ratio was taken to be among the values 0.0001, 0.001, 0.01, 0.1, and 0.999.

\begin{flushleft}
\begin{table*}
\centering \caption{SED fit. Input parameters of the CIGALE code.}
\label{SEDparam}
 \begin{tabular} {l l l }
\hline\hline
 \noalign{\smallskip}
Parameter   & &        \\
 \noalign{\smallskip}
\hline
 \noalign{\smallskip}
$\tau$ of old stellar population models (Gy)   & & 1,2,3,4,5,6,7     \\
Ages of old stellar population models (Gyr)     & & 13    \\
$\tau$ of young stellar population models (Myr) & &  5, 10, 20, 50, 100, 200    \\
Ages of young stellar population models (Myr)   & &  5, 10, 20, 50, 100, 200    \\
Fraction of young stellar population             & &  0.0001, 0.001, 0.01, 0.1, 0.999   \\
Metallicity                                      & &  0.02   \\
IMF                                              & &  Salpeter (1955)   \\
Stellar population libraries                     & &  Bruzual \& Charlot (2003)   \\
\noalign{\smallskip} \hline
\end{tabular}
\end{table*}
\end{flushleft}

The reduced $\chi^2$ provided by the best fit model equals 0.69 and 0.82 for NGC 5257 and NGC 5258, respectively. The results for the best fit  are shown in Table \ref{SED-fit}.
For NGC 5257, the averaged SFR over 100 Myr (42.3 M$_\odot \ yr^{-1}$) drops by almost a factor of two with respect to the averaged SFR over 10 Myr (27.2M$_\odot \ yr^{-1}$), but the latter does not differ from the instantaneous SFR (26.2 M$_\odot \ yr^{-1}$).
For NGC 5258, the variation between the averaged SFR over 100 Myr (27.7 M$_\odot \ yr^{-1}$) does not decrease significantly from the one computed over 10 Myr (22.9M$_\odot \ yr^{-1}$) and instantaneously (22.5 M$_\odot \ yr^{-1}$).
The stellar mass of the young population is more than ten times larger for NGC 5257 ( 2.63 $\times$ 10$^{8}$ M$_\odot$) than for NGC 5258 (2.22 $\times$ 10$^{7}$ M$_\odot$), while for the old population, masses are similar (1.024$\times$ 10$^{11}$ M$_\odot$ for NGC 5257 and 1.561 $\times$ 10$^{11}$ M$_\odot$ for NGC 5258). Both total stellar mass and total gas mass are 50\% larger for NGC 5258 ($M_\star$ = 1.561 $\times$ 10$^{11}$ M$_\odot$ and $M_{gas}$ = 6.43 $\times$ 10$^{10}$ M$_\odot$) than for NGC 5257 ($M_\star$ = 1.027 $\times$ 10$^{11}$ M$_\odot$ and $M_{gas}$ = 4.18 $\times$ 10$^{10}$ M$_\odot$).

\begin{flushleft}
\begin{table*}
\centering \caption{SED fit. Final Parameters}
\label{SED-fit}
 \begin{tabular} {l l l l l}
\hline\hline
 \noalign{\smallskip}
Parameter   & &  NGC 5257    & &  NGC 5258   \\
 \noalign{\smallskip}
\hline
 \noalign{\smallskip}
$\tau$ of old stellar population (Gyr)   & &  7   & & 7     \\
 Age of old stellar population\tablefootmark{*} (Gyr)      & & 13   & & 13   \\
$\tau$ of young stellar population (Myr) & & 100  & & 200    \\
Age of young stellar population (Myr)    & & 200  & & 400    \\
Fraction of young stellar population     & & 0.1  & & 0.1   \\
Instantaneous SFR (M$_\odot \ yr^{-1}$)             & & 26.17  & & 22.53       \\
Averaged SFR (over 10 Myr)   (M$_\odot \ yr^{-1}$)  & & 27.23  & &  22.93     \\
Averaged SFR (over 100 Myr)  (M$_\odot \ yr^{-1}$)  & & 42.33  & &  27.663     \\
Stellar mass of young population (M$_\odot$)       & &   2.63 $\times$ 10$^{8}$    & &  2.22 $\times$ 10$^{7}$     \\
Stellar mass of old population (M$_\odot$)         & &   1.024 $\times$ 10$^{11}$  & &   1.559 $\times$ 10$^{11}$     \\
Total stellar mass (M$_\odot$)    & &   1.027 $\times$ 10$^{11}$   & &  1.561 $\times$ 10$^{11}$     \\             
Total gas mass (M$_\odot$)        & &   4.18 $\times$ 10$^{10}$    & &    6.43 $\times$ 10$^{10}$     \\       
Metallicity\tablefootmark{*}                  & & 0.02     & &  0.02  \\
Reduced  $\chi^2$                & &  0.69    & &  0.82     \\
\noalign{\smallskip} \hline
\end{tabular}
\tablefoot{
\tablefoottext{*} Fixed values before running the code.}
\end{table*}
\end{flushleft}

\section{Discussion}
\label{disc}

The detailed study of the kinematics and dynamics of Arp 240 provides important information about the structure of each member of the pair, as well as about the stage of the encounter  as a whole. The velocity fields of NGC 5257 and NGC 5258 show velocity distributions following those of rotating discs with kinematic disturbances in localized regions. For NGC 5257, non-circular velocities traced by large velocity dispersion values (30 \ km \ s$^{-1} $ to 55 \ km \ s$^{-1} $ in top left panel in Fig. \ref{noncirc}) are associated with star-forming regions along the spiral arms, especially the large HII regions seen in the monochromatic image (bottom panel of Fig. \ref{arp240_img}). Large values are also seen on the tip of the eastern arm matching the beginning of the tidal tail.  According to  \citet{2010MNRAS.401.2113E}, the mean velocity dispersion of nearby isolated galaxies is $ \sim$  25 km \ s$^{-1}$. In this case, the large values in NGC 5257 could be due to the induced ongoing SF processes, as well as the gravitational effects of the interaction with the companion. Regarding the residual velocities map of this galaxy, high values are detected near the minor axis of the galaxy close to the centre of the galaxy, $\theta\sim90^\circ$ from the P.A. (bottom left panel of Fig. \ref{noncirc}).  In this region, low values of velocity dispersion are seen (top panel of Fig. \ref{noncirc}).   
Considering this angle, the tangential component of the velocity in Eq. 3 in Fuentes-Carrera et al. (2004) can be neglected,  indicating that an important fraction of non-circular velocities in this region lies in the radial direction on the plane of the galaxy, probably along the stellar bar.  For NGC 5258, large velocity dispersion values of the ionized gas match the location of the large HII region on the western arm of the galaxy (top panel of Fig. \ref{noncirc}). This is probably related to the intense star-forming processes in this region. Important residual velocity values are seen at the beginning of the tidal tail and the tidal bridge of this galaxy (bottom panel of Fig. \ref{noncirc}), which along with the large velocity dispersion values could be tracing the effect of the gravitational perturbations by the companion on both spiral arms of the galaxy. Comparing the velocity dispersion and residual velocity fields for both galaxies, we notice an increase of velocity dispersion at the location of the HII regions, while an excess of residual velocities is observed at the tip of the arms. The different locations of important velocity dispersion values and  residual velocity values is consistent with a scenario for which SF, shocks, and outflows increase closer to the disc and kinematical perturbations increase away from the disc in the spiral arms.

Globally, the RCs of both galaxies display a similar behaviour. The RCs remain symmetric up to a certain radius, then the rotation velocity on one side of the galaxy decreases, while for the other side, the velocity increases or remains constant. This type of bifurcation has been identified in simulations of major mergers \citep{2008A&A...484..299P} where the time evolution of these asymmetries has been studied in a pair of galaxies during the first passage. According to those authors, the rotation curve asymmetries appear right at the pericentre of the first passage and occur within a small time interval. Therefore the presence of bifurcation in the RC of each galaxy in the pair could be used as an indicator of the pericentre occurrence  and thus be related to the gravitational perturbation due to the presence of the equal-mass companion.

The pattern speed of NGC 5257 was derived using the TW method.  This is an important probe for estimating the impact of gravitational perturbations at the resonance radii of a galaxy, whether these perturbations are due to the internal structure of the galaxy or to interaction with an almost equal-size companion as in the present case. \citet{2003ApJ...599L..29C} showed the TW method can be used to determine multiple pattern speeds.  If more than one pattern is present, $\Omega_p$ will have contributions from the various patterns. In our case, the relatively large distance to the galaxies and the resulting spatial resolution of the velocity field did not allow us to do a detailed  $\Omega_p$ decomposition. We could, however, attempt to identify certain parts of the galaxy under the influence of certain density perturbation and derive the $\Omega_p$ considering only the average points falling in that region.  For NGC 5257,  $\Omega_p$ matches the velocity value of the ionized gas where the RC bifurcates at $R_{bif}=(17 \pm 2) \arcsec $. The fact that the velocity of the perturbation associated with the main disc of NGC 5257 matches the velocity of the ionized gas where the RC bifurcates indicates the extent of the effects of the gravitational perturbation by the massive companion galaxy on NGC 5257 versus the effects of the perturbations intrinsic to NGC 5257. Whatever density perturbation is present, its effect would be enhanced at this radius. We explored the possibility that $R_{bif}$, and thus $R_{cor}$, are related to the location of the SF processes in the galaxy by analysing the morphological and star-forming features at $R_{bif}$, that is at the position of the ellipse plotted on NGC 5257 in Figs. \ref{arp240_img},  \ref{arp240_ultracont}, \ref{arp240_bands}, and \ref{RC+morph}. The old stellar disc is confined by the ellipse,  while the UV maxima in the emission lies outside the ellipse.  The optical images indicate that the eastern spiral arm lies inside the ellipse. Outside the ellipse it appears to be tidally elongated to form the beginning of a tidal tail. Regarding the western arm of NGC 5257,  the brightest region in most wavelengths lies just outside the ellipse.

This pair of  galaxies is part of the sample studied by \citet{2007AJ....133..791S} in which most of the interacting galaxies display  nuclear and circumnuclear SF, implying that interactions can drive gas into the central region and trigger nuclear and circumnuclear SF before merging. However, in our case, both galaxies harbour all or most of the SF along their spiral arms. In particular for NGC 5257, the SF is confined to the spiral arms, close to the plotted ellipse.  For NGC 5258, the SF processes are confined to the SW region of the disc along the southern spiral arm, while some gas seems to have been fueled to the central parts of the galaxy (seen in the 4.5, 8.0, and 60 $\mu m$ images). 

Regarding the encounter,  the apparent bridge between the two galaxies is seen in the {\it F814W}, 4.5, and 8.0 $\mu m$ images. The bridge is broad and  joins both galaxies, at least in projection. It is also seen in the HI image by \citet{2005ApJS..158....1I};  however, it is not detected in any of the H$\alpha$ images. The tidal tail in NGC 5257 is a broad feature seen in UV, optical, NIR, 4.5, and 8.0 $\mu m$ images (Fig. {\ref{arp240_bands}).  It is not very extended, even in the HI image by \citet{2005ApJS..158....1I}. NGC 5257 does not look very perturbed, neither morphologically nor kinematically, by the presence of the massive companion, which could suggest this galaxy is suffering a retrograde encounter as shown by numerical simulations \citep{1972ApJ...178..623T,2004A&A...418L..27B,2010ApJ...725..353D}. The bifurcation observed in its RC seems to be associated with the beginning of the tidal arms.  On the contrary, the tidal tail of NGC 5258 is very broad and much more extended suggesting this galaxy  is experiencing a prograde encounter.  The relative orientation of the spin of the galaxies with the spin of the orbit  could explain why NGC 5258 is more perturbed than NGC 5257. According to simulations by \citet{2016MNRAS.459..720H}, NGC 5257 is experiencing a retrograde encounter, and the pair has already experienced the first peripassage which occurred $\sim$ 250 Myr ago. This time is in agreement with the findings by \citet{2008A&A...484..299P} where the bifurcation in the RC appears right after pericentre of the first passage and lasts $\sim$ 0.5 Gyr h$^{-1}$  Assuming the above orientation between both galaxies, the perturbation has been stronger for NGC 5258 and thus has experienced an induced SF episode before its companion, NGC 5257.  Indeed the SF in NGC 5257 is still confined to the outer parts of the galaxy, particularly the spiral arms, while it is no longer the case for NGC 5258. This delay in the triggering of the SF could explain the difference in the $\tau$ values in the SED fit for each galaxy.

The stellar mass ratio derived from the SED fit indicates that NGC 5258 is more massive than NGC 5257. This mass ratio is derived for each galaxy as a whole, while the RC decomposition shows that at small radii (0.37 $R_{25}$), NGC 5257 seems to be more massive than NGC 5258. This result is confirmed by the mass estimates using the \citet{1983A&A...125..394L} method. According to the simulations presented by \citet{2016MNRAS.459..720H}, Arp 240 is a 1:1 encounter, or a 1:2 encounter at most. Higher stellar mass at a large radius for NGC 5258 could imply that the SF on the outskirts of this galaxy has already happened and is currently  fueling the centre of the galaxy,  while NGC 5257 has  a more massive young stellar population confined to the outer parts of the galaxy. This can be thus considered as an $\sim$ 1:1 encounter where the parameters of the encounter have determined a  different SFH for each galaxy. Still both galaxies exhibit most SF far from their central parts.  

For the DM halo fit for these galaxies, the outer points of both RCs are affected by the interaction  in such a way that the DM halo fit is mainly associated with the inner parts of the galaxy. This corresponds to the part of the RCs before the bifurcation radii are observed.  Though this reduces the amount of information that can be derived from the fit, previous works \citep{1999AJ....118.2123B,2001AJ....121.1952B} have shown the importance of this part of the RC to infer the best DM halo model for individual spiral galaxies. In particular for Arp 240, it seems both RCs are best fit by a maximal disc and a pseudo-isothermal halo.

\section{Conclusions}
\label{ccl}

We have presented the kinematical and dynamical analysis of the interacting luminous infrared galaxy pair Arp 240 (NGC 5257/58). We used scanning Fabry-Perot interferometric H$\alpha$ observations to study the motions of different gas components. Through a morpho-kinematical analysis using direct images in different wavelengths, we have traced the location of recent and ongoing SF in the context of the interaction. In the case of NGC 5257, the SF is spread over the whole disc of the galaxy, while  for NGC 5258, the SF is mainly confined to a particular region in the southern arm where gas is being fueled towards the centre. Non-circular motions are detected in both galaxies. For both galaxies, in the inner parts of the discs, these motions are typically associated to spiral arms, and HII regions, and to streaming motion along the bar. For the outer parts of each galaxy, they are associated with the presence of the companion and the tidal response of the disc. 

The pattern speed of NGC 5257 was computed using the TW method. The corotation radius, $R_{cor}$, matches the location in the galaxy where the RC bifurcates. This radius could be pinpointing regions in the galaxy that are reacting to a tidal gravitational perturbation. The SF seems closely related to gravitational perturbation of the disc. This hypothesis will be explored through dynamical models of interacting galaxies in which the effects of perturbations can be traced.

This is a case where significant SF is confined to the arms of the galaxies. This does not mean gas will not be more centrally concentrated in the future. However, our study shows that important SF can occur in the outer parts of interacting galaxies. Our programme for the study of interacting galaxy pairs includes future numerical simulations that will consider gas and will take into account the extended kinematical information derived from the scanning FP interferometry observations in order to map the SF processes in both galaxies and the evolution of the perturbation during the encounter.

\begin{acknowledgements}
We thank the staff of the Observatorio Astron\'omico Nacional (OAN-SPM) for their support during PUMA data acquisition and C. Carignan for letting us use his mass model.
IFC thanks the financial support of IPN-SAPPI project 20181136 and CONACYT grant 133520. 
MR acknowledges financial support from grants DGAPA-PAPIIT (UNAM)  IN103116 and CONACYT CY-253085.
We acknowledge the use of the HyperLeda database (http://leda.univ-lyon1.fr) and the NASA/IPAC Extragalactic Database (NED).
Some of the data presented in this paper were obtained from the Mikulski Archive for Space Telescopes (MAST). STScI is operated by the Association of Universities for Research in Astronomy, Inc., under NASA contract NAS5-26555. 
This research has made use of the NASA/ IPAC Infrared Science Archive, which is operated by the Jet Propulsion Laboratory, California Institute of Technology, under contract with the National Aeronautics and Space Administration.
\end{acknowledgements}

\bibliographystyle{aa} 
\bibliography{arp240_VF} 

\end{document}